\documentclass[twocolumn]{aastex63}

\usepackage{natbib}
\usepackage{graphicx}
\usepackage{times}
\usepackage{apjfonts}
\usepackage{amsmath}
\usepackage{xcolor}
\usepackage{listings}
\usepackage{makecell}
\usepackage{verbatim}
\usepackage{enumitem}
\usepackage{natbib}
\setcitestyle{citesep={,}}

\providecommand\tablesize{\scriptsize}

\usepackage{soul}
\usepackage{amsmath}
\usepackage{amssymb}
\usepackage{xspace}
\usepackage{xifthen}
\usepackage{eso-pic}

\newcommand{\gold}{Y6 Gold\xspace}
\newcommand{\LCDM}{\ensuremath{\Lambda}CDM\xspace}

\newcommand{\nobjects}{691,483,608\xspace} 

\newcommand{\ngold}{669 million\xspace}
\newcommand{\nstars}{120 million\xspace}
\newcommand{\ngalaxies}{448 million\xspace}

\newcommand{\nflagszero}{635,487,439\xspace}

\newcommand{\areanimagesgrizy}{4913\xspace}

\newcommand{\footprintareaapp}{4923} 
\newcommand{\footprintarea}{4923.21} 

\newcommand{\astrorel}{27}

\newcommand{\photgaia}{1.8}

\newcommand{\maglimsnraperg}{24.7}
\newcommand{\maglimsnraperr}{24.4}
\newcommand{\maglimsnraperi}{23.8}
\newcommand{\maglimsnraperz}{23.1}
\newcommand{\maglimsnrapery}{21.7}

\newcommand{\maglimsofgapp}{24.2}
\newcommand{\maglimsofrapp}{23.9}
\newcommand{\maglimsofiapp}{23.4}
\newcommand{\maglimsofzapp}{22.7}
\newcommand{\maglimsofyapp}{21.3}

\newcommand{\maglimsofgup}{0.1}
\newcommand{\maglimsofrup}{0.1}
\newcommand{\maglimsofiup}{0.1}
\newcommand{\maglimsofzup}{0.1}
\newcommand{\maglimsofyup}{0.2}

\newcommand{\maglimsofglo}{0.2}
\newcommand{\maglimsofrlo}{0.2}
\newcommand{\maglimsofilo}{0.2}
\newcommand{\maglimsofzlo}{0.2}
\newcommand{\maglimsofylo}{0.2}

\newcommand{\magcompletebalrogg}{23.9} 
\newcommand{\magcompletebalrogr}{23.2}
\newcommand{\magcompletebalrogi}{22.7}
\newcommand{\magcompletebalrogz}{22.4}

\newcommand{\medfwhmg}{1.13} 
\newcommand{\medfwhmr}{0.99} 
\newcommand{\medfwhmi}{0.90} 
\newcommand{\medfwhmz}{0.87} 
\newcommand{\medfwhmy}{0.93} 

\newcommand{\starefficiency}{{94.6\%}\xspace}
\newcommand{\starcontamination}{{1.5\%}\xspace} 

\newcommand{\galefficiency}{{98.6\%}\xspace} 
\newcommand{\galcontamination}{{0.8\%}\xspace} 

\newcommand{\objdensity}{37.4}
\newcommand{\galdensity}{28.9}

\newcommand{\CHECK}[1]{{\textcolor{orange}{#1}}}

\mathchardef\mhyphen="2D

\newcommand{\roughly}{\ensuremath{ {\sim}\,} }

\newlength{\dhatheight}

\newcommand{\code}[1]{\texttt{#1}\xspace}

\newcommand{\unit}[1]{\ensuremath{\mathrm{\,#1}}\xspace}

\newcommand{\degree}{\ensuremath{{}^{\circ}}\xspace}

\newcommand{\amin}{\unit{arcmin}}
\newcommand{\asec}{\unit{arcsec}}

\newcommand{\magn}{\unit{mag}}
\newcommand{\mmag}{\unit{mmag}}
\newcommand{\e}{\unit{e^{-}}}

\newcommand{\secref}[1]{Section~\ref{sec:#1}}
\newcommand{\appref}[1]{Appendix~\ref{app:#1}}
\newcommand{\tabref}[1]{Table~\ref{tab:#1}}

\newcommand{\figref}[1]{Figure~\ref{fig:#1}}

\newcommand{\bandvar}[2][]{
  \ifthenelse{\isempty{#1}}{\var{#2}}{\var{#2\_#1}}
}

\newcommand{\magauto}[1][]{\bandvar[#1]{mag\_auto}}

\newcommand{\extmash}{\var{ext\_mash}}

\newcommand{\sofmash}{\var{sof\_mash}}

\newcommand{\flagsgold}{\var{flags\_gold}\xspace}
\newcommand{\flagsforeground}{\var{flags\_foreground}\xspace}
\newcommand{\flagsfootprint}{\var{flags\_footprint}\xspace}

\newcommand{\Gaia}{{\it Gaia}\xspace}
\newcommand{\ngmix}{\code{ngmix}}
\newcommand{\fitvd}{\code{fitvd}}
\newcommand{\bdf}{\code{BDF}}
\newcommand{\gap}{\code{GAp}}

\newcommand{\metadetect}{\code{metadetect}}
\newcommand{\BFD}{\code{BFD}}
\newcommand{\SWARP}{\code{SWarp}}
\newcommand{\swarp}{\SWARP}
\newcommand{\SCAMP}{\code{SCAMP}}
\newcommand{\scamp}{\SCAMP}
\newcommand{\SExtractor}{\code{SourceExtractor}}
\newcommand{\sextractor}{\SExtractor}
\newcommand{\PSFEx}{\code{PSFEx}}
\newcommand{\PIFF}{\code{PIFF}}

\newcommand{\HEALPix}{\code{HEALPix}}
\newcommand{\healpix}{\HEALPix}
\newcommand{\healsparse}{\code{healsparse}}
\newcommand{\nside}{\code{nside}}

\newcommand{\mangle}{\code{mangle}}
\newcommand{\decasu}{\code{decasu}}

\newcommand{\var}[1]{\ensuremath{\texttt{\MakeUppercase{#1}}}\xspace}

\newcommand{\SNR}{\ensuremath{\mathrm{S/N}}\xspace}

\newcommand{\xgboost}{\code{XGBoost}}

\providecommand\physrep{\ref@jnl{Phys.~Rep.}}
\providecommand\apjs{\ref@jnl{ApJS}}
\providecommand{\jcap}{\ref@jnl{JCAP}}

\newcommand{\drurl}{\url{https://des.ncsa.illinois.edu/releases}}

\shorttitle{DES \gold}
\shortauthors{DES Collaboration}

\begin{document}

\reportnum{DES-2023-0761}
\reportnum{FERMILAB-PUB-24-0932-PPD}

\title{Dark Energy Survey Year 6 Results: Photometric Data Set for Cosmology}


\author[0000-0001-8156-0429]{K.~Bechtol}
\affiliation{Physics Department, 2320 Chamberlin Hall, University of Wisconsin-Madison, 1150 University Avenue Madison, WI  53706-1390}
\author[0000-0002-1831-1953]{I.~Sevilla-Noarbe}
\affiliation{Centro de Investigaciones Energ\'eticas, Medioambientales y Tecnol\'ogicas (CIEMAT), Madrid, Spain}
\author[0000-0001-8251-933X]{A.~Drlica-Wagner}
\affiliation{Fermi National Accelerator Laboratory, P. O. Box 500, Batavia, IL 60510, USA}
\affiliation{Department of Astronomy and Astrophysics, University of Chicago, Chicago, IL 60637, USA}
\affiliation{Kavli Institute for Cosmological Physics, University of Chicago, Chicago, IL 60637, USA}
\author[0000-0002-9541-2678]{B.~Yanny}
\affiliation{Fermi National Accelerator Laboratory, P. O. Box 500, Batavia, IL 60510, USA}
\author[0000-0002-4588-6517]{R.~A.~Gruendl}
\affiliation{Center for Astrophysical Surveys, National Center for Supercomputing Applications, 1205 West Clark St., Urbana, IL 61801, USA}
\affiliation{Department of Astronomy, University of Illinois at Urbana-Champaign, 1002 W. Green Street, Urbana, IL 61801, USA}
\author[0000-0001-9194-0441]{E.~Sheldon}
\affiliation{Brookhaven National Laboratory, Bldg 510, Upton, NY 11973, USA}
\author[0000-0001-9376-3135]{E.~S.~Rykoff}
\affiliation{Kavli Institute for Particle Astrophysics \& Cosmology, P. O. Box 2450, Stanford University, Stanford, CA 94305, USA}
\affiliation{SLAC National Accelerator Laboratory, Menlo Park, CA 94025, USA}
\author[0000-0001-8318-6813]{J.~De~Vicente}
\affiliation{Centro de Investigaciones Energ\'eticas, Medioambientales y Tecnol\'ogicas (CIEMAT), Madrid, Spain}
\author[0000-0002-6904-359X]{M.~Adamow}
\affiliation{Center for Astrophysical Surveys, National Center for Supercomputing Applications, 1205 West Clark St., Urbana, IL 61801, USA}
\author[0000-0003-3312-909X]{D.~Anbajagane}
\affiliation{Kavli Institute for Cosmological Physics, University of Chicago, Chicago, IL 60637, USA}
% ...

\collaboration{106}{(DES Collaboration)}

\email{K.~Bechtol (kbechtol@wisc.edu), I.~Sevilla (ignacio.sevilla@ciemat.es), A.~Drlica-Wagner (kadrlica@fnal.gov)}

\begin{abstract}

We describe the photometric data set assembled from the full six years of observations by the Dark Energy Survey (DES) in support of static-sky cosmology analyses. 
DES \gold is a curated data set derived from DES Data Release 2 (DR2) that incorporates improved measurement, photometric calibration, object classification and value added information. 
\gold comprises nearly $5000\,\deg^2$ of $grizY$ imaging in the south Galactic cap and includes \ngold objects with a depth of $i_{AB} \sim 23.4$\,mag at $\SNR \sim 10$ for extended objects and a top-of-the-atmosphere photometric uniformity $< 2 \mmag$. 
\gold augments DES DR2 with simultaneous fits to multi-epoch photometry for more robust galaxy shapes, colors, and photometric redshift estimates.
\gold features improved morphological star-galaxy classification with efficiency \galefficiency and contamination \galcontamination for galaxies with $17.5 < i_{AB} < 22.5$.
Additionally, it includes per-object quality information, and accompanying maps of the footprint coverage, masked regions, imaging depth, survey conditions, and astrophysical foregrounds that are used for cosmology analyses.
After quality selections, benchmark samples contain \ngalaxies galaxies and \nstars stars.
This paper will be complemented by online data access and documentation.

\end{abstract}

\section{Introduction}\label{sec:intro}

Optical and near-infrared imaging surveys have played an essential role in developing the standard model of cosmology that invokes a cosmological constant and cold, collisionless dark matter (\LCDM).
For a fixed allocation of telescope observing time, broadband photometric surveys assemble the largest samples of galaxies that can be used for statistical analyses, while also providing the opportunity to combine several complementary probes of the cosmic expansion history and growth of structure \citep[e.g.,][]{des_2019_multi,kids_2021_multi}.
The current generation of imaging surveys, such as the Pan-STARRS1 surveys \citep[PS1;][]{panstarsurveys}, the Hyper Suprime-Cam Subaru Strategic Program \citep[HSC-SSP;][]{HSCPDR2}, the Ultraviolet Near-Infrared Optical Northern Survey (UNIONS),\footnote{\url{https://www.skysurvey.cc/}} the Kilo-Degree Survey \citep[KiDS;][]{kids_2012_dr4}, the DESI Legacy Imaging Surveys \citep{Dey:2019}, the DECam Local Volume Exploration survey \citep[DELVE;][]{Drlica-Wagner:2021}, and the Dark Energy Survey \citep[DES;][]{DES:2005,DES:2016ktf} collectively provide deep, multi-band imaging over nearly the entire high-Galactic-latitude sky, and have cataloged more than a billion galaxies and thousands of supernovae spanning 10 billion years of cosmic history.
Together with spectroscopic surveys like  eBOSS \citep{eboss_2020_cosmology} and DESI \citep{2016arXiv161100036D}, imaging surveys yield measurements of the expansion rate and large-scale structure in the late-time universe that are complementary to precision measurements of the early Universe \citep[e.g.,][]{planck_cosmology}. 
Combined analyses of the early- and late-time observations rigorously test the \LCDM paradigm, with percent-level measurement uncertainties on the \LCDM model parameters \citep[e.g.,][]{DESY3KP, DESY5SN, 2024arXiv240403000D}.

To support both cosmological and other astronomical investigations, DES data products are publicly released via two pathways, as summarized in \tabref{releases}.
Two general DES data releases, DES DR1 \citep{DESDR1} and DES DR2 \citep{DESDR2} provide coadded images and associated object catalogs for the first three and six years of the survey, respectively.
Further image-processing algorithms, survey characterization, and value-added data products have been developed to control systematic uncertainties at the level required for static-sky cosmology analyses by the DES Collaboration.
These data products and validation analyses have been compiled into DES ``Gold'' releases, including SVA1 Gold, Y1 Gold \citep{Y1Gold}, and Y3 Gold \citep{Y3Gold}.
Here, we present the final iteration of the DES Gold data products, the Year 6 (Y6) Gold, assembled from the full six-year DES data set and intended to support legacy cosmology analyses using the DES data.

DES \gold is based on the same DECam data that were released as DES DR2.
As expected, \gold is very similar in depth and extent to DR2, but provides additional photometry measurements from multi-epoch fitting, photometric redshift estimates, footprint and foreground masks, additional summary flags, survey property maps and an improved object classification scheme.
These products are described in the sections that follow, and \tabref{summary} summarizes a set of metrics that describe the DES \gold release.

The structure of the paper is as follows. In \secref{overview}, we provide an overview of DES and the resultant six-year data set. \secref{processing} provides a brief summary of the DES data processing with a focus on new algorithms implemented for \gold. The value-added content of the \gold catalog is described in \secref{gold}, whereas the new ancillary maps are detailed in \secref{maps}. We remark on known issues in \secref{issues} and mechanisms for using the data in \secref{using}. We conclude by discussing the importance of \gold in \secref{conclusions}.

\begin{deluxetable*}{c c c c c c c c}
\tablewidth{0pt}
\tabletypesize{\tablesize}
\tablecaption{Dark Energy Survey data releases \label{tab:releases}}
\tablehead{
\colhead{Release} & \colhead{Area} & \colhead{Depth} & \colhead{Objects} &  \colhead{Photometry uniformity} & \colhead{Supplemental data} & \colhead{Reference}
\\
\colhead{} & \colhead{(sq.deg.)} & \colhead{($i$ band)} & \colhead{} &  \colhead{(mmag)} & \colhead{}
}
\startdata
SVA1 Gold & $\sim 250$ & 23.68 & 25M & $< 15$ & Photo-$z$ &  \\
Y1 Gold & 1786 & 23.29 & 137M & $< 15$ & BPZ/DNF photo-$z$, MOF, maps, classification& \cite{Y1Gold} \\
DES DR1 & 5186 & 23.33 & 399M & $< 3$ & \makecell{None} & \cite{DESDR1} \\
Y3 Gold & 4946 & 23.34 & 388M &  $< 3$ & \makecell{BPZ/DNF photo-$z$, SOF/MOF, maps, classification} & \cite{Y3Gold} \\
Y3 Deep Fields & 5.88 & 25.0 & 2.8M (1.6M NIR) & $ <5$ & \makecell{$ugrizYJHK$ bands} & \citealt*{y3deepfields} \\
DES DR2 & \areanimagesgrizy & 23.8 & 691M &  $\sim 2$ & None & \cite{DESDR2} \\
\gold & $\footprintareaapp$ & \maglimsnraperi & 669M &  $< 2$ & \makecell{DNF photo-$z$, \fitvd/\gap, maps, classification}  & This work \\
\enddata
\tablecomments{All releases are publicly accessible at \url{https://des.ncsa.illinois.edu/releases}. Quoted depth corresponds to ${\rm S/N} = 10$ in $2\asec$ diameter apertures. SOF and MOF are multi-epoch pipelines replaced by \fitvd, described in Section \ref{sec:fitvd}. The \gold area is computed for simultaneous two-exposure coverage in $griz$, whereas the DES DR2 area is quoted for one exposure in all five $grizY$ bands.}
\end{deluxetable*}

\begin{deluxetable*}{l c c c c c c}
\tablewidth{0pt}
\tabletypesize{\tablesize}
\tablecaption{Key numbers and data quality summary for the DES Wide Survey (\gold; this work). All magnitudes are in the AB system. 
\label{tab:summary}}
\tablehead{
\colhead{Parameter} & \multicolumn{5}{c}{Band} \\
 & $g$ & $r$ & $i$ & $z$ & $Y$
}
\startdata
\multicolumn{6}{c}{Wide Survey (this work)} \\
\hline
Median PSF FWHM ($\asec$) & $\medfwhmg$ & $\medfwhmr$ & $\medfwhmi$ & $\medfwhmz$ & $\medfwhmy$ \\
Sky Coverage ($griz$ intersection, deg$^{2}$) & \multicolumn{5}{c}{$\footprintareaapp$} \\ 
Coadd Median Astrometric Relative Precision (angular distance, mas) & \multicolumn{5}{c}{\astrorel} \\ 
Photometric Uniformity vs. Gaia (mmag) \tablenotemark{{\tiny a}} & \multicolumn{4}{c}{$\photgaia$} & \nodata \\ 
Median Coadd Magnitude Limit, $1.95 \asec$ diameter ($\SNR = 10$) & $\maglimsnraperg$ & $\maglimsnraperr$ & $\maglimsnraperi$ & $\maglimsnraperz$ & $\maglimsnrapery$ \\ 
Coadd $90\%$ Completeness Limit for extended objects (mag) \tablenotemark{{\tiny b}} & $\magcompletebalrogg$ & $\magcompletebalrogr$ & $\magcompletebalrogi$ & $\magcompletebalrogz$ & \nodata \\ 
Multi-Epoch Galaxy Magnitude Limit ($\SNR = 10$, BDF) \tablenotemark{{\tiny c}} & $\maglimsofgapp^{+\maglimsofgup}_{-\maglimsofglo}$ & $\maglimsofrapp^{+\maglimsofrup}_{-\maglimsofrlo}$ & $\maglimsofiapp^{+\maglimsofiup}_{-\maglimsofilo}$ & $\maglimsofzapp^{+\maglimsofzup}_{-\maglimsofzlo}$ & $\maglimsofyapp^{+\maglimsofyup}_{-\maglimsofylo}$ \\ 
Galaxy Selection ($17.5 \leq \magauto[i] \leq 22.5$; $\extmash = 4$) & \multicolumn{5}{c}{Efficiency \galefficiency; Contamination \galcontamination} \\
Stellar Selection~ ($17.5 \leq \magauto[i] \leq 22.5$; $0 \leq \extmash \leq 1$) & \multicolumn{5}{c}{Efficiency \starefficiency; Contamination \starcontamination} \\
Object density ($\amin^{-2}$) \tablenotemark{{\tiny d}}& \multicolumn{5}{c}{Overall: \objdensity; Galaxies: \galdensity} \\
\enddata
\tablenotetext{a}{Photometric uniformity measured vs.\ Gaia's $G$ band, which encompasses DECam's $griz$.}
\tablenotetext{b}{As measured by \textsc{Balrog} (Anbajagane \& Tabbutt et al., in prep.).}
\tablenotetext{c}{Median values with $16\%$ and $84\%$ percentile errors from the magnitude limit distribution.}
\tablenotetext{d}{Object density determined for all objects in \gold footprint outside foreground regions, and the subset of those classified as high-confidence galaxies (\sofmash >= 3).} 
\end{deluxetable*}

\section{Survey overview and derived data sets}
\label{sec:overview}

\subsection{Survey overview}

DES used the 570 megapixel Dark Energy Camera \citep[DECam;][]{decam} on the 4-m Blanco Telescope at Cerro Tololo Interamerican Observatory in Chile to image the southern Galactic cap in five broad-band filters ($grizY$) extending from $\sim 400$\,nm to $\sim 1060$\,nm \citep{DESDR2}. 
Images were collected on 760 distinct full or half nights between 2013 August 15 and 2019 January 9. DES operated in two survey modes \citep{2019arXiv191206254N}:

\begin{itemize}
\item The \textbf{Wide-Field Survey} is optimized for cosmological analyses using weak gravitational lensing, galaxy clustering, and galaxy clusters.
The Wide-Field Survey spans $\sim5000\,\deg^2$ that was imaged with dithered tilings in $grizY$ (see \citealt{DESDR2} for details).
The Wide-Field Survey footprint was designed to significantly overlap with the South Pole Telescope survey \citep{SPTref} and SDSS Stripe 82 \citep{2009ApJS..182..543A}, and it includes a connection region to enhance overall calibration.
The Wide-Field Survey constitutes the basis for the \gold data set. 

\item The \textbf{Supernova Survey} is a time-domain survey of 10 DECam fields, amounting to a total of $\sim27\,\deg^2$ that was imaged in $griz$ with an approximately weekly cadence \citep{2020AJ....160..267S} with minimal dithering imaging. Difference imaging analysis of the Supernovae Survey fields has enabled the discovery of thousands of Type Ia supernovae (SNIa) and precision photometric lightcurves are computed following \citet{2019ApJ...874..106B}. 
\end{itemize}

\subsection{Derived data sets}

The DES Collaboration has assembled several high-level data products derived from DECam imaging collected by DES:

\begin{itemize}
\item \textbf{DES DR2} and \textbf{\gold} are assembled from data collected by the Wide-Field Survey. A total of 72,217 DECam exposures were deemed of sufficient quality to pass on to the next step of image detrending, calibration and finally coaddition and object detection.  DES DR2 and \gold contain the same number of objects, which were detected and measured by a pipeline based on the \sextractor \citep{sextractor} software (see \secref{processing} and \citealt{DESDR2} for details). Additional pipelines are run over the DES DR2 coadded catalogs to obtain the \gold data set, which are the focus of this paper. The \gold catalog caters to several science cases including extragalactic astronomy, galaxy cluster cosmology, and cosmology analyses using the large scale distribution of the positions of the objects according to their photometric properties.

\item The \textbf{shear catalogs} are specialized data sets used for applications that involve weak gravitational lensing measurements. In Year 6, two different shear catalogs were produced using data from the Wide-Field Survey: the Bayesian Fourier Domain method \citep[\BFD;][]{BFD} catalog uses the sames detections as DR2 and \gold, whereas the \metadetect \citep{2023OJAp....6E..17S} pipeline produces a set of 5 distinct catalogs based on the same images, which are coadded in a parallel pipeline (Yamamoto, Becker et al. in prep.) and have 5 different sets of detections. The \metadetect catalogs are produced to calibrate bias in shear measurements produced by noise, modeling errors and selection (including detection), as described in \citet{2023OJAp....6E..17S}. The shear catalogs will be released separately.

\item Finally, the Supernova Survey exposures are coadded to produce the \textbf{Deep Field data sets}. Together with DECam imaging of the COSMOS field,\footnote{\url{http://cosmos.astro.caltech.edu}} this specialized processing enables high S/N measurements of galaxies ${\sim}1.5$ to 2.0\,\magn fainter than the Wide-Field Survey. A subset of these data have been combined with deep near-infrared imaging to produce a reference object catalog used for various applications in DES cosmology analyses (\citealt*{y3deepfields}, \citealt{2024A&A...686A..38T}).  A larger region of DECam and NIR deep fields are being processed and analyzed (Gruendl et al., in prep).
\end{itemize}

\begin{figure*}
\includegraphics[width=\textwidth]{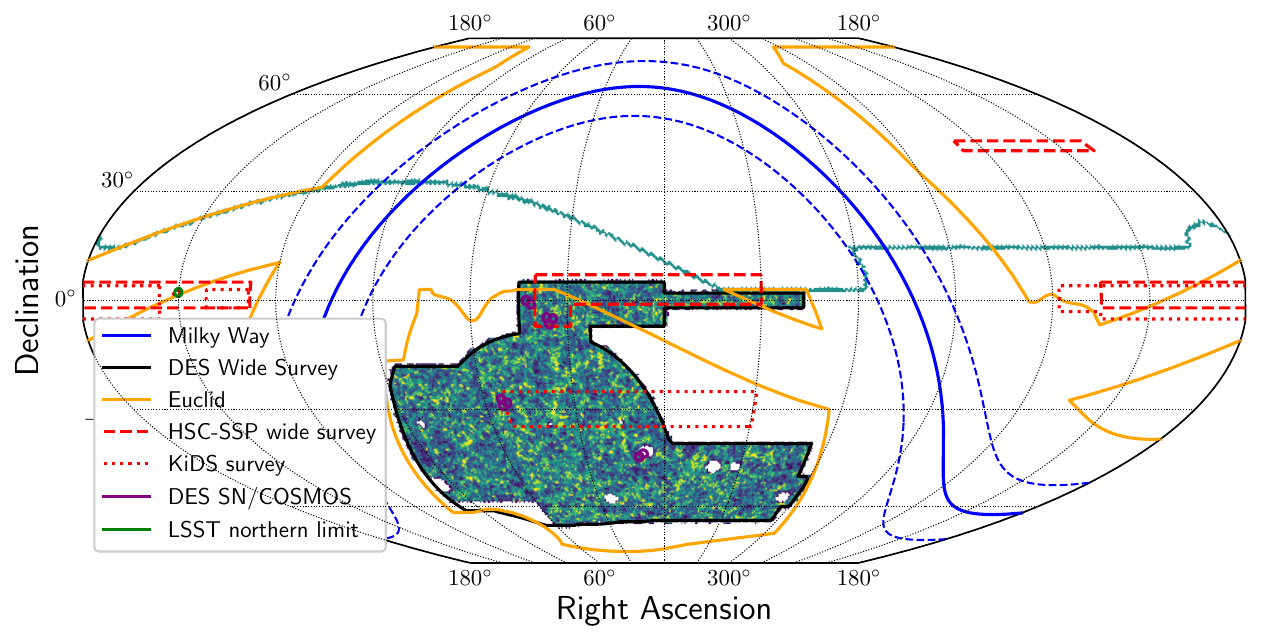}
\caption{\label{fig:footprint}
DES footprint in equatorial coordinates. 
The $\roughly 5000 \deg^2$ wide-area survey footprint is shown as a black outline, with overplotted \textsc{redMaGiC} \citep{redmagic} galaxies, for the redshift bin $z = [0.5, 0.6]$. The supernova field locations are also shown as purple circles with their approximate area to scale (corresponding to one full DECam field of view). A few other footprints from present and future major photometric surveys are shown for reference. This and the other skymap plots included in this work use the equal-area McBryde-Thomas flat-polar quartic projection \citep{McBryde:1949}.
}
\end{figure*}

\figref{footprint} shows the DES footprint, including the Wide-Field Survey and Supernovae Survey.  
Given the cosmological goals of the survey, DES avoids the Galactic plane to minimize stellar foregrounds and extinction from interstellar dust.

In this work, all quoted data quality characteristics (e.g., \tabref{summary}) refer to the subset of exposures included in the DES DR2 coadded images unless stated otherwise.

\subsection{Comparison to Y3 Gold}

Given the emphasis on the use of \gold for cosmology measurements, it is worth highlighting improvements over the previous cosmology release, Y3 Gold:

\begin{itemize}
    \item Greater depth and uniformity (as shown in Table \ref{tab:releases}) so that $70\%$ more objects were detected with respect to the previous release.
    \item Improved photometry as a consequence of the above.
    \item Better point photometric redshift precision ($\sim20\%$ improvement) and accuracy at $z \sim 1$.
    \item A more robust star-galaxy classification, that includes an additional purity level for galaxy samples, as well as a new boosted decision-tree-based algorithm that improves the performance over a larger range of magnitudes.
    \item Additional flagging of foreground objects and artifacts. 
    \item Footprint and foreground maps with higher resolution, and a new map to correct for Galactic cirrus.
\end{itemize}

\section{Data Processing}
\label{sec:processing}

The DES Data Management system (DESDM; \citealt{desdm}), running at the National Center for Supercomputer Applications (NCSA) as the core data processing center, converted raw DECam data to detrended and coadded images and catalogs. 
These data were distributed to the DES Collaboration in the form of files and database tables. 
Additional value-added columns and ancillary data products were produced across several of the collaborating DES institutions and collected at NCSA for distribution.

\subsection{Detrending}

The single-exposure (or ``single-epoch'') detrending of instrumental signatures for DES DR1 is described in \citet{desdm}. 
The main processing changes implemented for DES DR2 (and shared by \gold) are described in \citet{DESDR2} and summarized here:

\begin{itemize}
\item New calibrations (biases, darks, flats, etc.) were derived for the later survey years (\secref{calibration}). 
\item The astrometric reference catalog was updated to Gaia DR2 \citep{2016A&A...595A...2G,2018A&A...616A..14G}.
\item A utility was introduced to search for and mask a region of anomalous charge arising from a variable hot pixel (``light bulb'') on CCD 46, which was first noticed in exposures shortly after 2017 September 1.
\item A utility was implemented to search for and mask occurrences of an amplifier instability that arose for amplifier B of CCD 41 starting 2018 August 15.  The instability manifests intermittently as a charge transfer inefficiency that results in streaked row reads with varying background charge. The utility looks for images where the amplifier background is discontinuous from one row to the next and flags the entire amplifier when triggered.\footnote{These utilities are publicly available among the DESDM software repositories: \url{https://github.com/DarkEnergySurvey/pixcorrect/}}
\item A utility was added to use the detections of streaks on individual CCDs (such as those created by satellites) to identify potential trails on adjacent CCDs.
\item The single-epoch catalog effective detection threshold was lowered to ${\rm S/N} \gtrsim 3$ due to configuration changes in \PSFEx \citep{psfex}, as well as changes to the detection threshold and deblending settings for the initial {\sextractor}-generated catalogs.
\item The sky subtraction algorithm, described in sections 3.2 and 4.3 of \citet{desdm} remains unchanged from Y3: a simultaneous fit to all 60 or 61 CCDs of an exposure with a low order PCA template is performed to remove large scale scattered light gradients and a pupil ghost.  During image coaddition, a constant median background level is subtracted from each CCD to bring the sky level close to zero.
\end{itemize}

\subsection{Calibration}
\label{sec:calibration}
\gold measurements build upon the astrometric and photometric calibration of DR2. 
Thus, the underlying \gold astrometric positions, tied to Gaia DR2 \citep{2016A&A...595A...2G,2018A&A...616A..14G}, and photometric calibrations, derived by the forward global calibration module \citep[FGCM;][]{FGCM}, are the same as those provided in the DR2 public data release.

The median astrometric precision of the coadd averaged over the DES footprint is estimated to be 27 mas \citep{DESDR2}.
Based on a comparison between Gaia $G$-band synthesized magnitudes transformed from stellar DES $griz$ magnitudes ($G_{\rm pred}$) and measured Gaia $G$-band magnitudes ($G_{\rm meas}$) from Gaia DR3  \citep{gaiadr3}, the photometric uniformity of \gold is  at the 1.8 mmag (0.18\%) RMS level or better  across the DES footprint \citep{rykoff_2023_des_standard_stars}.
The absolute calibration of the catalog is computed with reference to the Hubble Space Telescope CalSpec standard star C26202. 
Including systematic errors, the absolute flux system is known at the $\approx 1\%$ level. 
\citet{rykoff_2023_des_standard_stars} present DES $grizY$ magnitudes for 17 million stars with $i$-band magnitudes mostly in the range $16 \lesssim i \lesssim 21$ as a photometric calibration reference catalog for optical imaging in the southern hemisphere.

All photometry for \gold is based on the \code{APER8} system used for FGCM, with aperture corrections computed as described in \secref{aperture_corrections}. Briefly, the \code{APER8} system normalizes PSF-fitted stellar photometry to aperture photometry within a fixed radius (the 8th in a set of 12 apertures) of diameter 5.84 arcsec.

\begin{figure*}
\includegraphics[width=\textwidth]{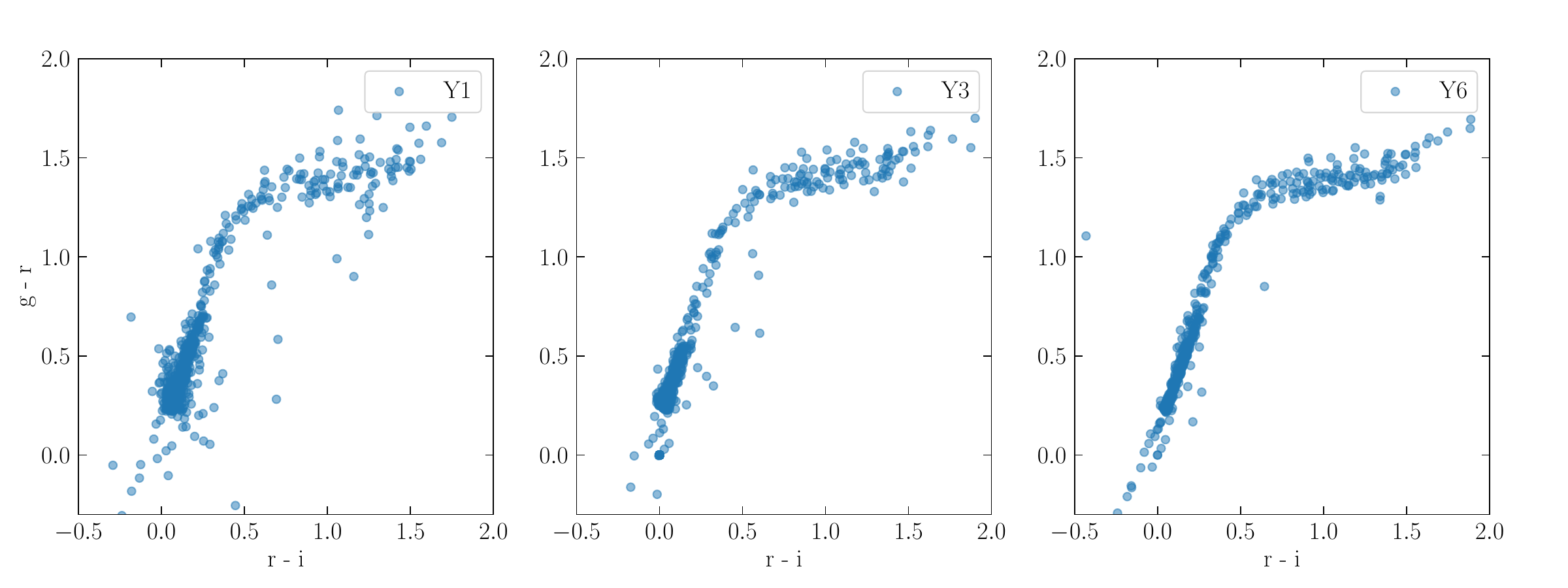}
\caption{\label{fig:slr_showcase}Illustration of the increasing calibration and selection quality in successive data releases, from Y1 Gold (\emph{left}), Y3 Gold (\emph{center}) to \gold (\emph{right}). 
This figure shows the stellar locus for selected stars in the outskirts of the globular cluster NGC\,1261.
}
\end{figure*}

In Figure \ref{fig:slr_showcase} a qualitative illustration of the improvement of the stellar locus in the vicinity of the globular cluster NGC\,1261 is shown in successive Gold releases. The tightness of the stellar locus is indicative of the superior photometric quality in Y6.

\subsection{Coaddition and Object Detection}

Object detection is performed on combined $r+i+z$ coadd detection images (the three bands simultaneously) using \sextractor with an approximate threshold of ${\rm S/N} \gtrsim 5$, resulting in a set of \nobjects objects across the survey footprint, identical to that released in DES DR2.
For the DES Y6 cosmological analyses, we run additional forced photometry measurement pipelines starting from this initial detection catalog and compile the results in \gold.

To facilitate multi-epoch photometry, we use the DR2 catalog to build Multi-Epoch Data Structures \citep[MEDS;][]{2016MNRAS.460.2245J} comprised of ``postage stamp'' images  extracted from the coadd and single-epoch images for each object. 
Along with the science frame data, the postage stamps also carry the weight, mask, background, and PSF model information. 
Two types of MEDS files were created for Y6.
The first set carries single-epoch PSF models from \PSFEx, and was used for the Y6 Gold photometry measurements using \fitvd (see Section \ref{sec:fitvd}). 
The second set incorporates PSF models generated by \PIFF \citep{piff}, and was used with the \code{BFD} shear measurement pipeline.
The other shear pipeline, \code{metadetect}, also uses \PIFF PSF models, but uses cell-based coadds rather than MEDS files \citep{2020ApJ...902..138S}.

\subsection{Multi-Epoch Photometry}
\label{sec:fitvd}

The previous Y1 and Y3 Gold science pipelines developed a multi-object, multi-epoch, multi-band fit (MOF) for each object, where objects were grouped according to a friends-of-friends (FoF) algorithm \citep{Y1Gold, Y3Gold}. A single-object-fit (SOF) variant  was also employed that masked nearby objects rather than performing a simultaneous multi-object fit.  
For DES Y6, object deblending was first performed on coadded images using a new code (\code{shredder}, see \citealt*{y3deepfields}).  A multi-band, multi-epoch fit was then performed on the original single-epoch images, with neighbors subtracted using the models from the deblender. The new framework, \fitvd (first described in \citealt*{y3deepfields}) performed the neighbor subtraction and ran fitting algorithms from \ngmix \citep{2014MNRAS.444L..25S}, using postage stamp images stored in MEDS files.  Again a simpler version using only masking of neighbors was also performed.  Ultimately we used only the version with masked neighbors for most analyses, because the photometry was quite consistent for the two techniques.   Our interpretation is that blending is not a significant effect on photometry for most objects in the DES data, which is significantly shallower than that of HSC-SSP and the forthcoming Rubin Observatory's Legacy Survey of Space and Time (LSST).

The basic \fitvd algorithm solves for position, flux, intrinsic size and ellipticity parameters. For DES Y6, the fit is performed assuming two different source models. The first was a simple, zero-size point-source model to provide PSF photometry. The second assumed a Bulge+Disk Fixed (\bdf) model (the size ratio of bulge effective radius to disk effective radius is fixed to unity). In each case, the object model is convolved with a parametric model of the PSF consisting of a five-component Gaussian mixture model fit to the reconstructed \PSFEx model for each CCD. 
The fit is then performed simultaneously on all single-epoch observations of a given object across all bands.

The DES Y6 \bdf model is fit with the following free parameters:
\begin{itemize}
    \item The flux in each band
    \item The offset from the fiducial position in arcseconds
    \item The size squared ($T = \langle x^2 \rangle + \langle y^2 \rangle$).
    \item The 2-component ellipticity ($\{\code{g1,\,g2}\}$).
    \item The fraction of the flux in a bulge (DeVaucouleurs S\'ersic model with index $n=4$), with the remaining flux assigned to an exponential disk (S\'ersic model with index $n=1$) (\code{frac\_dev}).
\end{itemize}
The only free parameters of the PSF model are the flux and the offset from the fiducial position.

The \gold catalog includes measured fluxes, magnitudes, and uncertainties for the PSF and \bdf models calculated in each of the individual $grizY$ bands. For the \bdf model, we include the color covariance terms corresponding to each of the off-diagonal elements in the variance matrix of $grizY$ photometry. In addition, the \gold catalog includes the positional offset, size, ellipticity, and \code{frac\_dev}, along with their associated uncertainties, from the \bdf model fit.

In addition to PSF and \bdf models, Gaussian aperture (\gap) fluxes are also calculated \citep{2022ApJS..258...15E, y3deepfields}.  A Gaussian with ${\rm FWHM} = 4$\,arcsec is multiplied with the \bdf model and integrated analytically to give a total flux.  While this will necessarily undercount the flux, it is less sensitive to the wings of the object model.  This is useful when the \code{frac\_dev} of the object is very noisy, which can result in very noisy total flux estimates.  Artificially high values of \code{frac\_dev} can result in large total fluxes due to the large outer extent of the DeVaucouleurs profile.  Using \gap fluxes can be more useful, for example, when trying to estimate the photometric transfer function with a source injection scheme such as \code{Balrog} (\citealt{2022ApJS..258...15E}, 
 Anbajagane \& Tabbutt et al., in prep.).

Additional multi-epoch processing was performed to derive a morphological object classifier based on the difference of Gaussian-weighted fluxes.
Fluxes were derived using a Gaussian weighted with a size equal to the size of the PSF and a Gaussian with a weight function increased by 5\%.
We define the ratio of these measurements, $C_{\rm raw} = -2.5 \log_{10}(F_{\rm dil}/F)$. 
For point-like objects, $C_{\rm raw}$  will be equal to a particular value that depends on the PSF.
This value is calibrated by performing the same Gaussian fit using the PSF model, $C_{\rm psf} = -2.5 \log_{10}(F_{\rm dil, psf}/F_{\rm psf})$. 
We then define the concentration parameter, $\code{CONC} = C_{\rm psf} - C_{\rm raw}$.
Point-like objects occupy a narrow locus at $\code{CONC} = 0$, while galaxies generally have $\code{CONC} > 0$.
While the \code{CONC} parameter was not found to perform significantly better for star/galaxy classification than a prescription based on the \bdf fits (\secref{classification}), the multi-epoch Gaussian fit is very robust and measurements of \code{CONC} exist for nearly all objects in the \gold catalog (in contrast, the \bdf fits fail for $\roughly 0.2\%$ of objects).

\section{\gold Object Catalog}
\label{sec:gold}

The DES \gold catalog is a merger of columns drawn from DES DR2, the multi-epoch photometric measurements described in the previous section, and new quantities described in this section. These include corrections to the measured multi-epoch photometry, star-galaxy classification, object quality flags that summarize other flags from measurement pipelines and features of the data, and a photometric redshift estimator.

\subsection{Photometric Corrections}
\label{sec:photometric_corrections}


\gold includes columns for PSF and \bdf model photometry that have been normalized to the \var{mag\_aper\_8} system used for global photometric calibration, and dereddened with a fiducial interstellar extinction correction. These are labeled as \var{\_CORRECTED} in the catalogs and are the most uniform across the survey footprint.

\subsubsection{Aperture Corrections}
\label{sec:aperture_corrections}

The DES DR2 photometric zeropoints were established by FGCM using the PSF flux measurements of stars in the single-epoch catalogs measured by \sextractor assuming the \PSFEx model. These measurements are placed on the \var{APER\_8} (5.84-arcsec-diameter aperture) system through the normalization of the \PSFEx model.
An aperture correction is necessary to place the multi-epoch \fitvd PSF and \bdf model measurements of stars on this same photometric system. 
This aperture correction is estimated at the location of each catalog object using the \ngmix Gaussian mixture model fit to the PSF (\secref{fitvd}) and calculating the fraction of the PSF model flux contained within a 5.84-arcsec-diameter circular aperture (\var{APER\_8}) relative to the total PSF model flux.
This correction is typically $\sim 2\%$, in the sense that the corrected flux in the \var{APER\_8} system for stars is $\sim 2\%$ fainter than the uncorrected total PSF flux. 
This procedure (i.e., assuming a point-like object) is used to calculate an aperture flux correction for all objects in the \gold catalog. The correction is reasonably accurate for very small galaxies (i.e., galaxies with angular sizes of $< 1$\,arcsec  before convolution with the PSF). However, larger galaxies have a larger fraction of their light extending beyond the \var{APER\_8} aperture, and thus the aperture corrections estimated assuming a point-like source does not reduce the flux enough to make the \fitvd model measurements match the \var{APER\_8} measurements. Note that this aperture correction is not intended to address long-standing issues concerning how to deal with extended galaxy profiles in the calculation of total galaxy magnitudes, but rather simply ensures that the \fitvd PSF and \bdf magnitudes of stellar objects are on the FGCM system, and that the flux measurements converge in the small-size limit of unresolved galaxies.

\subsubsection{Interstellar Extinction}
\label{sec:reddening}

The \gold table includes a column containing the $E(B-V)$ values from the reddening map of \citet{sfd98} (SFD98) extracted at the location of each catalog object. 
The $E(B-V)$ values were obtained using a linear interpolation of the Zenithal Equal Area projected map distributed by SFD98 and are the same as provided in DES DR2. 
The FGCM magnitudes can be corrected by an amount $A_b = E(B-V) \times R_b$, where $R_b$ is computed per band as described in \citet{DESDR1} using the DES Standard Bandpasses. 
Following Section 4.2 of \citet{DESDR1}, we incorporate a renormalization of the original SFD98 reddening map \citep[$N = 0.78$;][]{2011ApJ...737..103S} to our fiducial reddening coefficients so that these coefficients can be used directly with $E(B-V)$ values from the SFD98 (see \secref{extinction} for details on this renormalization).
The values for $R_b$ are thus the same as for DES DR1 and DR2, and are replicated here for completeness: $R_g = 3.186$, $R_r = 2.140$, $R_i = 1.569$, $R_z = 1.196$, and $R_Y = 1.048$.

\subsection{Object Classification}
\label{sec:classification}

\begin{figure*}[t!]
\includegraphics[width=0.49\textwidth]{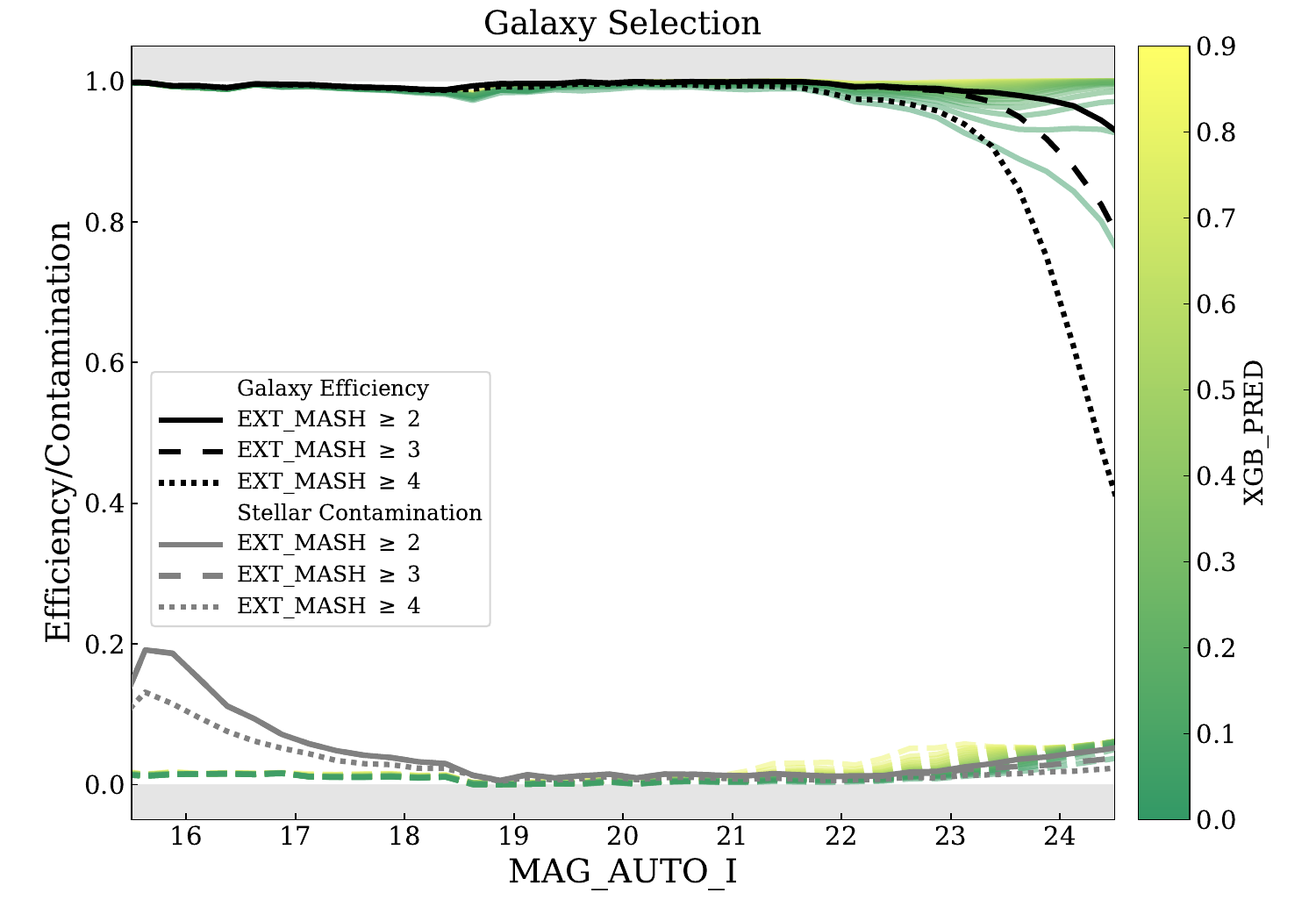}
\includegraphics[width=0.49\textwidth]{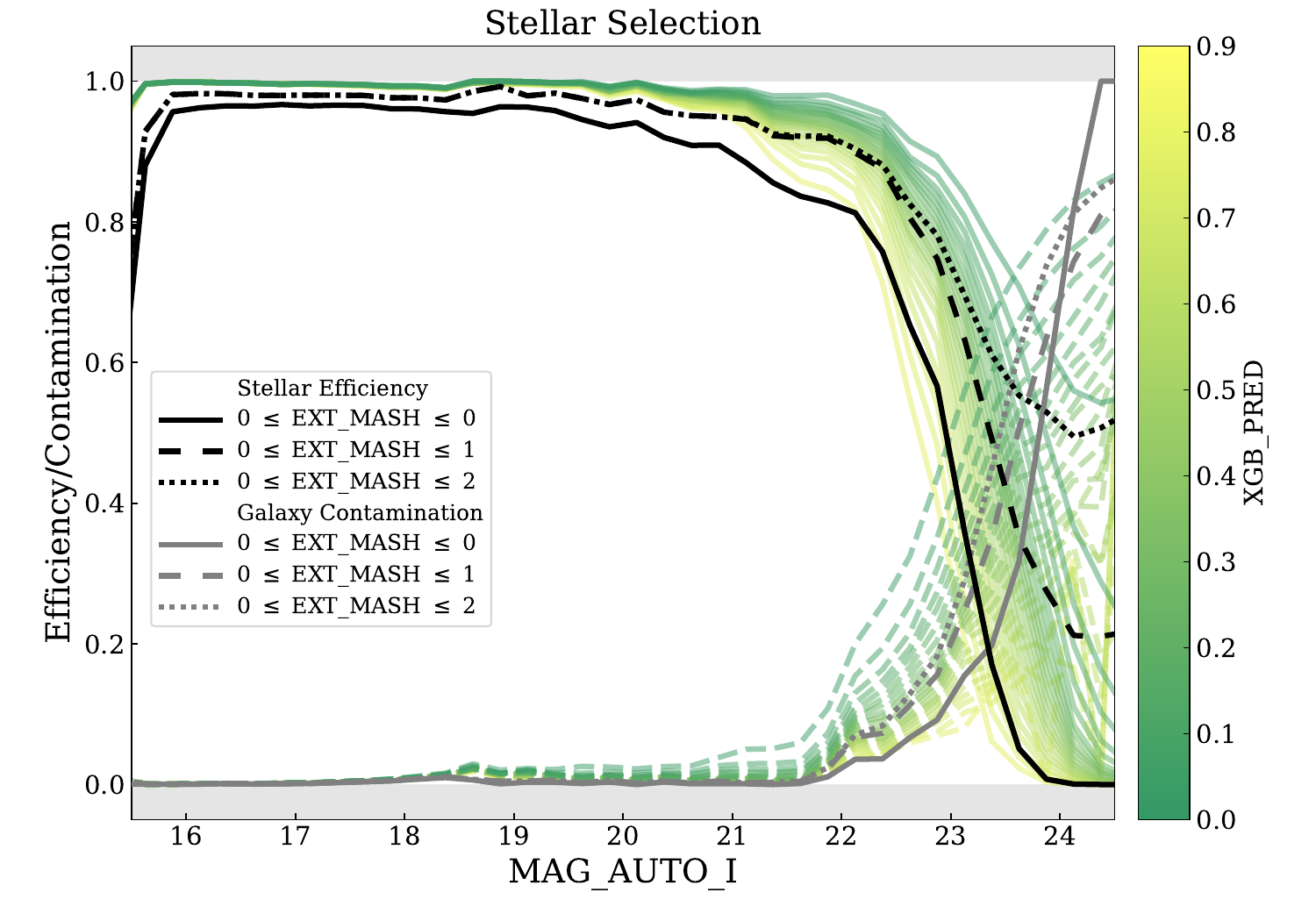}
\caption{Performance of morphological star/galaxy classification in DES \gold. The conventional cut-based classifier is shown in solid, dashed, and dotted lines for three different object selections. Black lines correspond to selection efficiency (true positive rate), while gray lines show contamination (false discovery rate). The green/yellow colored lines show the performance of the \xgboost classifier for different selections on the continuously valued XGB\_PRED classifier output. \label{fig:stargal}}
\end{figure*}

Following previous DES analyses \citep[e.g.,][]{Y3Gold, DESDR2}, we define high-quality samples of point-like objects (e.g., stars, quasars) and extended objects (e.g., galaxies) based on morphological measurements.
We assign each object in the \gold catalog to a morphological class based on the multi-epoch measurements, the weighted average of the single-epoch measurements, and measurements on the coadded images with larger values corresponding to higher-confidence extended objects (\var{EXT\_MASH}; \tabref{extclass}).
Furthermore, we train a gradient boosted decision tree algorithm \citep[\xgboost;][]{Chen:2016} on the multi-epoch and single-epoch weighted average measurements to automate the classification processes (\var{EXT\_XGB} and \var{XGB\_PRED}).
Here, we summarize the performance of these morphological classifiers, while more details can be found in \appref{classification}.  

\begin{deluxetable}{c c }
\tablewidth{\columnwidth}
\tablecaption{Morphological object classes \label{tab:extclass}}
\tablehead{ 
\colhead{\hspace{1cm} Value \hspace{1cm}} & \colhead{\hspace{1cm} Description } 
}
\startdata
\hspace{1cm} 4  & \hspace{1cm} Ultra-pure galaxy sample \\
\hspace{1cm} 3  & \hspace{1cm} High-confidence galaxies \\
\hspace{1cm} 2  & \hspace{1cm} Mostly galaxies \\
\hspace{1cm} 1  & \hspace{1cm} Likely stars \\
\hspace{1cm} 0  & \hspace{1cm} High-confidence stars \\
\hspace{1cm} $-$9 & \hspace{1cm} Data not available \\
\enddata
\tablecomments{Discrete object classes assigned in the \code{EXT\_MASH}, \code{EXT\_FITVD}, \code{EXT\_XGB} variables. The \code{EXT\_COADD} and \code{EXT\_WAVG} variables do not include class 4.}
\end{deluxetable}

We assess the efficiency (true positive rate) and contamination (false discovery rate) of each of our output classes in the bright ($i < 18.5$\,mag) and faint ($i > 18.5$\,mag) domains using a high-Galactic-latitude region of the footprint. 
In the bright domain, we use infrared data from the Vista Hemisphere Survey \citep[VHS DR5;][]{McMahon:2013} to classify stars and galaxies as demonstrated in \citet{Baldry:2010} and \citet{Sevilla-Noarbe:2018}. 
We perform a 0.5 arcsec astrometric match between the \gold and VHS catalogs at $0\degree < \alpha_{2000} < 45\degree$ and $\delta_{2000} \sim 0\degree$ ($-68\degree < b < -48\degree$), and define a stellar vs.\ non-stellar classification based on DES $(g - i)$ optical color versus VHS $(J - K_s)$  infrared color.
In the faint domain, we use data from HSC-SSP PDR3 deep/ultra-deep SDSX field \citep{HSCPDR3}, which is located at $\alpha_{2000}, \delta_{2000} \sim 35.8\degree, -4.6\degree$ ($b \sim -58.5\degree$) and has superior depth and image resolution ($0\farcs75$ median).
We perform a 0.5 arcsec astrometric match between the \gold and HSC-SSP catalogs, and define stellar vs.\ non-stellar classifications based on the HSC-SSP concentration parameter defined as the difference between the \code{i\_psfflux\_mag} and the \code{i\_cmodel\_mag} following the prescription described in \citet{DESDR1} and \citet{Drlica-Wagner:2021}.

\figref{stargal} shows the efficiency (true positive rate) and contamination (false discovery rate) of the DES \gold \code{EXT\_MASH} and \xgboost classifiers compared to the VHS-based color classification at the bright end ($i < 18.5$\,mag) and HSC-SSP PDR3 morphological classification at the faint end ($i > 18.5$\,mag).
Black and gray lines show the efficiency and contamination respectively for the conventional cut-based \code{EXT\_MASH} classifier, with different line styles corresponding to different object classes, as detailed in the legend of \figref{stargal}.
In addition, the green/yellow lines show the efficiency/contamination of a broad range of object classes that can be defined based on the continuous \xgboost predictor output, \code{XGB\_PRED}. 
Similar to DES Y3 Gold, we see an increase in the stellar contamination in the bright galaxy sample at $\code{MAG\_AUTO\_I} < 19$\,mag.
This increase in contamination is partially driven by the increasing fraction of stars in the object sample at bright magnitudes, in addition to contamination from double stars that are morphologically extended but classified as stars based on their infrared colors.
Interestingly, we find that the morphological \xgboost classifier is much more robust to this contamination at the bright end, suggesting that these contaminants occupy a distinct region of the morphological parameter space, which can be identified through more complex machine-learning techniques.  The \gold ultra-pure galaxy class ($\var{EXT\_MASH} = 4$) achieves 90\% completeness for galaxies at a magnitude 0.5~mag fainter than Y3 Gold does, with no increase in contamination. Nonetheless, cosmological analyses that require a very pure sample of galaxies may consider using additional color cuts to remove bright contaminants from the galaxy sample (see for example Weaverdyck et al. in prep.).

\appref{classification} discusses the relative performance of these classifiers, with \tabref{extclass_performance} providing a summary.

\subsection{Object Quality Flags}
\label{sec:flagsgold}

The \flagsgold column is a bitmask used to identify objects that present unusual features in the measurement process or that are deemed unphysical. 
Flagged objects can be excluded as appropriate for a given analysis using bitwise operations. The \flagsgold bits and the number of affected objects can be found in \tabref{flagsgold}.
Several object quality flag categories are defined to indicate objects with potentially suspect measurements.
The approach taken in \gold is similar to that applied in Y3 Gold, and dominantly focuses on the \fitvd (formerly MOF/SOF) and \sextractor processing flags. 
Several additional bits have been added to flag other types of spurious objects and objects with compromised photometry, which we describe below.
We expect that most science applications will select objects with $\flagsgold = 0$, which includes \nflagszero objects in the full catalog.

\begin{deluxetable*}{l c l}
\centering
\tablewidth{0pt}
\tabletypesize{\scriptsize}
\tablecaption{Summary of bitmask values and warning descriptions for the \flagsgold column \label{tab:flagsgold}}
\tablehead{
  \colhead{Bit} & \colhead{Number of objects affected} & \colhead{Description}
}
\startdata
1 & 1388296 (0.2\%) & \fitvd $\var{FLAGS} != 0$, Indicates problems in \fitvd processing \\
2 & 833130 (0.1\%) &  	Any \SExtractor \var{FLAGS\_\{GRIZ\}} $>$ 3, Standard \SExtractor quality selection \\
4 & 2830733 (0.4\%) & Any of \SExtractor $\var{IMAFLAGS\_ISO\_\{GRIZ\}} != 0$, Saturated objects \\
8 & 11444931 (1.7\%) & Super-spreader objects; $\var{BDF\_T} * \var{BDF\_T\_ERR} > 30$ and $\var{BDF\_T} / \var{BDF\_T\_ERR} < 3$ \\
16 & 41483901 (6\%) & Possible noise objects; any of $riz$ magnitudes approx below the S/N $\approx 1$ threshold (as determined by exposure time calculations) \\
32 & 8627937 (1.2\%) & Extreme color outliers; $r-i$ or $i-z$ colors outside the $[-5,5]$ range \\
64 & 240448 (0.03\%) & Phantom objects; objects with $\var{NEPOCHS}=0$ for all bands and $\var{MU\_EFF\_MODEL} > 26$ for $riz$ bands, but with $\var{MAG\_AUTO\_I} < 22$ \\
\enddata
\end{deluxetable*}

A new class of flagged objects with suspect measurements in \gold, so-called ``super-spreader'' objects, are identified as having extremely large sizes and large size measurement uncertainty when measured by \fitvd.
These objects were initially identified in Y3 synthetic source injection analyses \citep[see Figure 21 of][]{2022ApJS..258...15E} and subsequently in the Y6 redMaGiC sample as outliers possessing much too large photometric uncertainty relative to their brightness \citep[see][for a description of the redMaGiC sample]{redmagic}.
We attribute these fitting failures to catastrophic over-estimation of the object size that occurs more frequently in crowded fields and regions with structured diffuse light (\appref{cirrus}).

We flag another class of ``noise objects'' that are sufficiently faint to have a predicted signal-to-noise ratio smaller than unity. 
To identify these objects, we defined flux thresholds using the DECam exposure time calculator assuming a lunar phase of 10 days after new Moon and the approximate integrated exposure time of the DES coadds (900\,seconds). 
Objects in the DES Wide-Field Survey with measured magnitudes fainter than the AB magnitude thresholds of $\{r,i,z\} = \{26.5,26.2,25.6\}$\,mag are flagged as likely noise artifacts. Note that this selection may remove $r$-band outliers which may be of interest for certain analyses.

Extreme color outliers are defined to catch lingering reflections and residual noise artifacts that would hinder photo-$z$ estimates. This flag is found to be very correlated with the noise objects. For similar reasons, we use the $riz$ detection bands and define color outliers as any of the colors $r-i$ or $i-z$ to fall outside the $[-5,5]$ range, using the AUTO magnitudes.

Phantom objects are objects that have a bright magnitude in coadds, but have not been detected in individual single epochs. These are coming mostly from diffraction spikes and excess light around bright stars.

\subsection{Photometric Redshifts}
\label{sec:photoz}

\begin{figure*}
\includegraphics[width=0.49\textwidth]{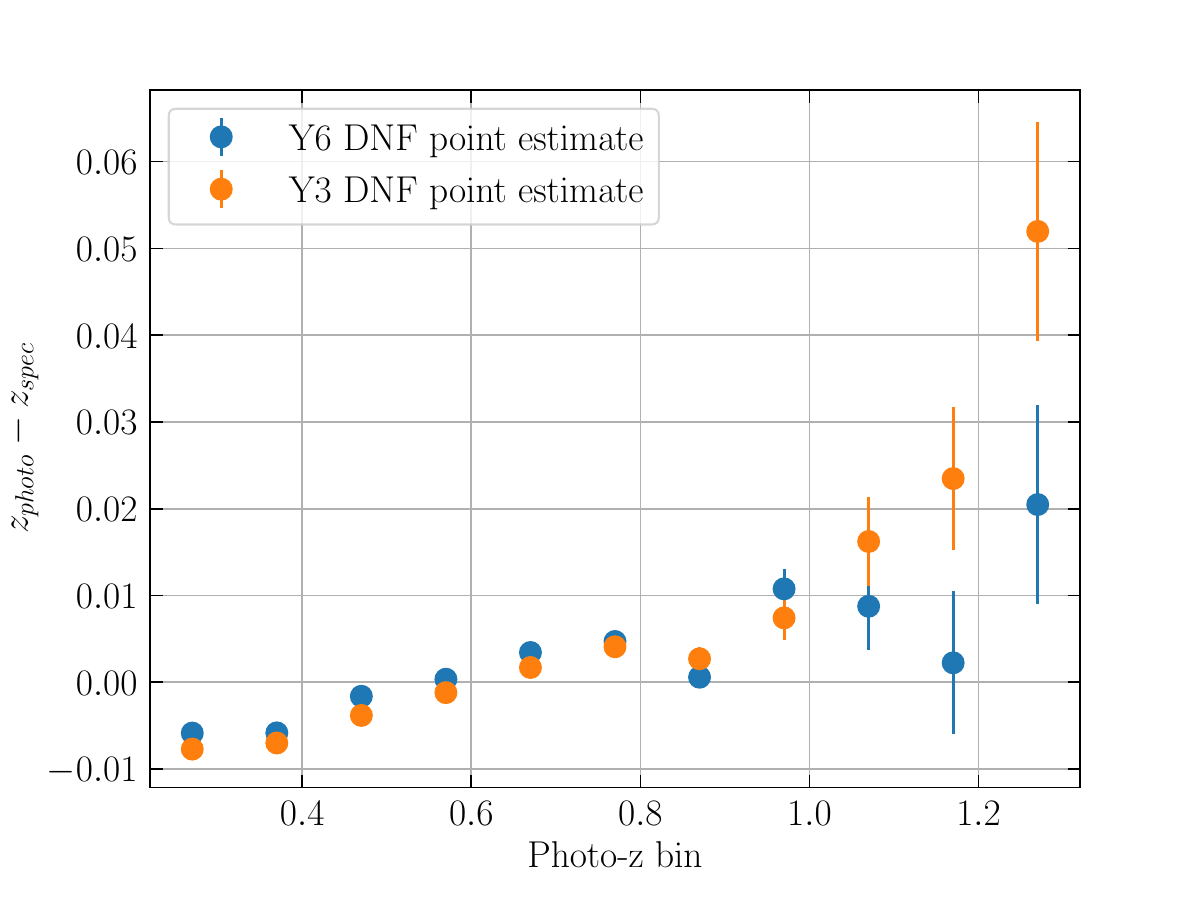}
\includegraphics[width=0.49\textwidth]{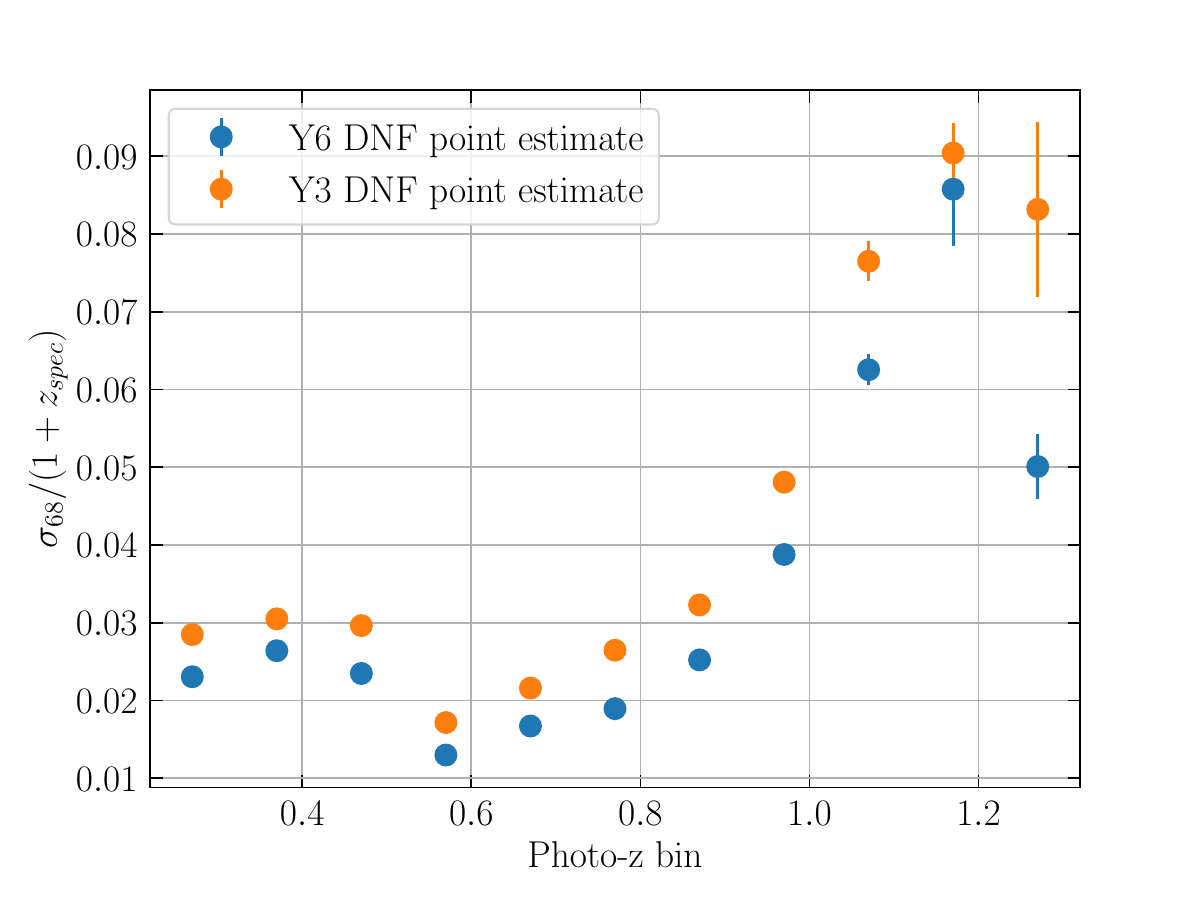}
\caption{(Left) Photo-$z$ bias for the default DNF estimate (\code{DNF\_Z}) using a spectroscopic reference sample. (Right) Photo-$z$ scatter measured as the 68\% containment value for the default \code{DNF\_Z} estimate using a spectroscopic reference sample.}
\label{fig:dnf_pz}
\end{figure*}

The \gold catalog provides default photometric redshift estimates for every object based on their \fitvd-corrected (by extinction and aperture)  magnitudes and colors. These estimates are based on the directional neighbourhood fitting \citep[DNF;][]{dnf} code, that was successfully applied to and validated on the DES Y3 data \citep{Y3Gold, 2024A&A...686A..38T}. 
It uses a nearest-neighbor approach with a directional metric that accounts simultaneously for magnitudes and colors to obtain the photometric redshift for each object (\code{DNF\_Z}) using around 80 neighbors (\var{DNF\_NNEIGHBORS}). The photo-$z$ error (\var{DNF\_ZSIGMA}) is estimated as the quadratic mean of the uncertainty due to photometric errors (\var{DNF\_ZERR\_PARAM}) and the uncertainty obtained from the residuals of the fit (\var{DNF\_ZERR\_FIT}). On the other hand, only the single nearest neighbor is used to provide a point value (\var{DNF\_ZN}) to construct an $N(z)$ estimate. Other approaches relying on calibration with deep fields \citep[e.g.,][]{2021MNRAS.505.4249M, 2024MNRAS.527.2010G} may be used additionally for these distributions, and will be released in the corresponding analyses. 

\figref{dnf_pz} shows some standard photo-$z$ metrics for a Y3 and \gold selection matched to a spectroscopic data sample. This data set has been compiled with the framework described in \citet{2018A&C....25...58G} and includes 545,796 spectroscopic redshifts deemed of optimal quality (\var{FLAG\_DES} = 4, according to the classification in \citealt{2018A&C....25...58G}). Note that the spectroscopic sample is brighter than \gold, and further studies are required to understand its performance in detail \citep[e.g.,][]{2020MNRAS.496.4769H}. The lens sample for the combined weak lensing and galaxy clustering analyses, as well as the galaxy sample used for measurements of baryon acoustic oscillations, are drawn from intermediate depth samples of \gold ($i \lesssim 22.5$\,mag) for which these metrics would approximately apply. Most of the relevant galaxies (i.e., galaxies in the magnitude range of interest) from the spectroscopic reference dataset come from public releases such as SDSS DR16 ($62\%$, \citealt{SDSSDR16}) or 2dF ($9\%$, \citealt{2dFDR}), with deeper and narrower surveys covering the fainter magnitudes such as DEEP2 ($2\%$, \citealt{deep2DR4}) and VVDS ($1\%$, \citealt{2013A&A...559A..14L}). VIPERS also constitutes a major source of moderate to faint galaxies with a very broad color coverage ($5\%$, \citealt{vipers}). DES specific proprietary programs were done using AAOmega on the AAT to complement these spectra ($3\%$, \CHECK{\citealt{2020MNRAS.496...19L}}). Finally, an assortment of other shallow and deep datasets were included as well, for increased color and redshift coverage.

\section{Ancillary Maps}
\label{sec:maps}

\begin{figure*}[]
\includegraphics[width=\textwidth]{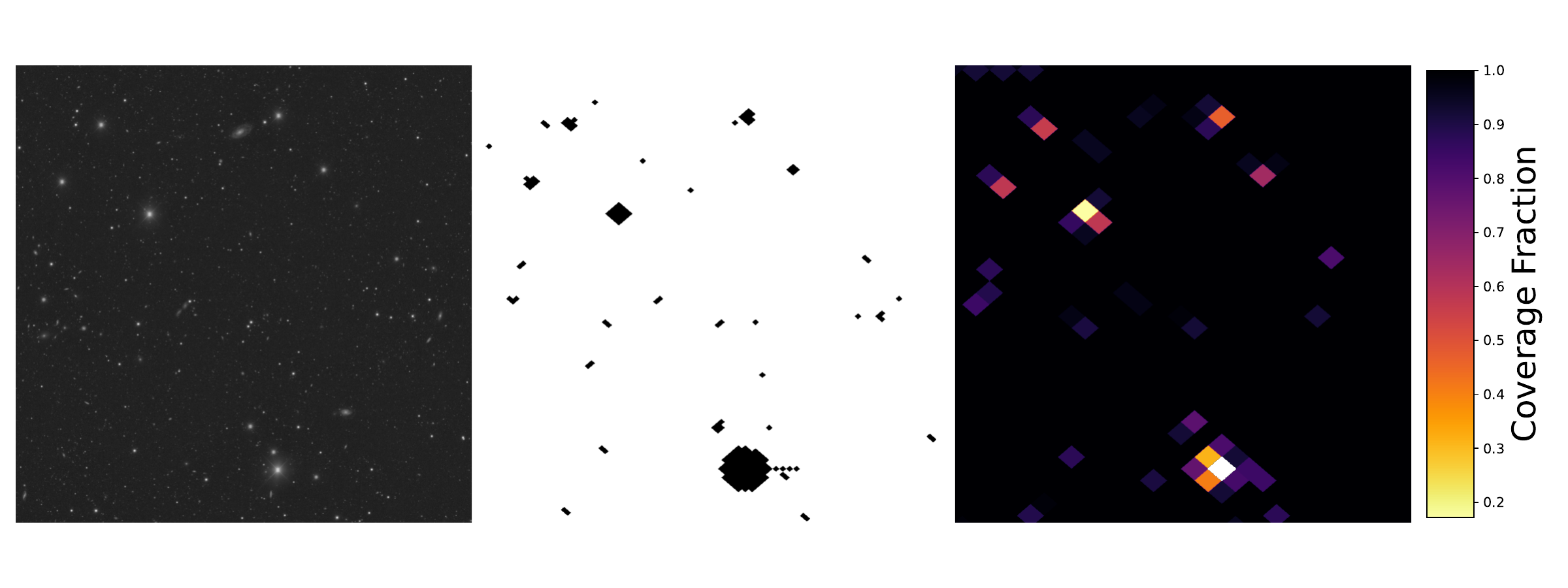}
\caption{A side-by-side comparison of the different footprint resolutions using tile DES0219$-$0541 in the $g$ band as an example. The coadd image is shown as a reference on the left. In the middle panel, the binary mask at the highest resolution available in \gold is shown ($\nside=16384$). Finally, the coarser resolution ($\nside=4096$) mask with approximate coverage fraction is shown on the right. This coverage fraction represents how many high resolution subpixels contain valid information for this band.
}
\label{fig:healsparse_resolution}
\end{figure*}

In addition to the object catalog, the \gold data set includes several maps of the survey geometry, survey properties, and astrophysical foregrounds that complement the interpretation of the catalogs.
A technical advance introduced in \gold is the use of the \healsparse\footnote{\url{https://healsparse.readthedocs.io}} software to store the map and mask data. \healsparse is a sparse implementation of \healpix in Python that optimizes memory usage by using a coarser resolution in those areas of the sky which are not covered. In the case of \gold, this allows practical usage of maps with \healpix resolution of \nside = 16384 corresponding to pixels of area 0.046 square arcminutes.
The enhanced resolution is sufficient for detailed representation of the gaps between sensors on the DECam focal plane mosaic as well as masked regions around individual bright stars  (\figref{healsparse_resolution}).

The maps described in this section include the survey footprint (\secref{footprint}), a mask of astrophysical foregrounds (\secref{foregrounds}), survey property maps (\secref{properties}), and a mask of diffuse foregrounds (i.e., Galactic cirrus and nebulosity; \appref{cirrus}). 

\subsection{Footprint}
\label{sec:footprint}

The \gold footprint is a description of the angular mask that contains the regions of the sky that are deemed useful for cosmological analyses with the \gold catalog. The source of `truth' for which parts of the sky have been observed are \mangle \citep{MollyMangle} files that describe in detail what is the geometry of the CCDs on overlapping exposures for each band. In turn, these complex, high resolution maps are represented into a standardized \healpix map of \nside = 16384 resolution, which represents the underlying survey geometry at suitable accuracy for our needs. The \gold footprint then is represented via a binary \healsparse file of \nside = 16384 resolution, 
constructed by applying certain conditions on the \mangle maps at these resolutions. These include:

\begin{itemize} 
    \item At least 2 exposures in each of $g$, $r$, $i$, and $z$ in the \code{NUM\_IMAGE} survey property map.
    \item $f_{griz} > 0.5$, where $f_{griz}$ is the fraction of each pixel that has simultaneous coverage in the four bands when considering \nside = 4096 pixels, which is used for compatibility purposes as some Y3 analyses used this coarser version.
\end{itemize}

A footprint map with only the two-exposure condition is also made available. For the case of the coarse \nside = 4096 footprint version, a complementary map with the same resolution is available denoting the fraction of each pixel that has simultaneous coverage in the four bands  ($0 \leq f_{griz} \leq 1$). 

Each object in \gold has an associated \var{FLAGS\_FOOTPRINT} value which is equal to the footprint map value at the position of the object (using the full-resolution \var{ALPHAWIN\_J2000}, \var{DELTAWIN\_J2000} coordinates) provided that the object also has the \sextractor quantities $\var{NITER\_MODEL\_\{G,R,I,Z\}} > 0$ for every band. This ensures that the object indeed has the observations in all four bands, in case it happens to be in one of the residual regions of a valid footprint pixel, lacking some of the observations in the key bands.
The total \gold footprint area using the high-resolution map with the conditions described above is $\footprintarea\,\deg^2$.
The \gold area computed in this way is slightly larger than the DES DR2 area (4913\,deg$^2$), which was estimated requiring at least one exposure in each of the five $grizY$ bands.

The \gold footprint area is slightly smaller than the Y3 Gold footprint area, $4945.87\,\deg^2$. This is a result of the increase of both the survey depth and the threshold for minimum number of exposures (changed from 1 to 2) between the two releases.

For a threshold of 2 exposures in each of $g$, $r$, $i$, and $z$, the equivalent footprint area for Y3 Gold is $4495.26\,\deg^2$, demonstrating the increase in coverage from Y3 to \gold.

\subsection{Foreground Mask} 
\label{sec:foregrounds}

\gold includes a mask to identify regions of the footprint that are likely to be impacted by the presence of bright astrophysical foreground objects.
Similar to Y3 Gold, we define the foreground mask for bright stars, globular clusters, and nearby galaxies. 
For bright stars, we mask regions from a magnitude-dependent radius that was derived from the density of DES Y6 objects where the $i$-band magnitude measured by the bulge-disk fit is much brighter than that \sextractor \texttt{AUTO} measurement (i.e., $\var{BDF\_MAG\_I}-\var{MAG\_AUTO\_I}<-1$).

This mask is constructed as a \healpix bit map ($\nside=4096$), and catalog objects are assigned a \var{FLAGS\_FOREGROUND} value corresponding to the sum of the bits in that position of the map, according to their sky coordinates ($\alpha_{2000}, \delta_{2000}$). 
In cases where the mask radius is smaller than a single \healpix pixel, the pixel containing the object is used as the mask. The specific bits in the foreground mask are defined in \tabref{foregrounds}. A \healsparse high resolution map ($\nside=16384$) is also made available upon release, though \flagsforeground in the table does not follow this convention.

\begin{itemize}
    \item \textbf{Bit 1, Gaia bright stars:} bright stars from Gaia\ DR2 ($G < 7$) were masked based on the $G$-band magnitude.
    \item \textbf{Bit 2, Yale bright star catalog:} stars from the Yale Bright Star Catalog \citep{Hoffleit:1991} were masked based on the $V$-band magnitude.
    \item \textbf{Bit 4, 2MASS bright stars:} bright stars ($4 < J < 8$) magnitude from the 2MASS catalog \citep{2006AJ....131.1163S} were masked based on the $J$-band magnitude.
    \item \textbf{Bit 8, Gaia moderately bright stars:} moderately bright stars ($7 < G < 11.5$) from Gaia\ DR2 were masked based on the $G$-band magnitude.
    \item \textbf{Bit 16, 2MASS moderately bright stars:} moderately bright ($8 < J <12$) stars from the 2MASS catalog \citep{2006AJ....131.1163S}.
    \item \textbf{Bit 32, Bright galaxies:} area around large, nearby galaxies found in the HyperLEDA\footnote{\url{http://leda.univ-lyon1.fr/}} catalog \citep{hyperleda}.
    \item \textbf{Bit 64, Milky Way satellites:} Milky Way globular clusters and classical dwarf spheroidal galaxies in the footprint were masked (\tabref{satellites}). 
    \item \textbf{Bit 128, Large Magellanic Cloud (LMC) periphery:} the periphery of the LMC ($60 < \alpha_{2000} < 100 \deg$ and $-70 < \delta_{2000} <  -58 \deg$) was masked due to the significant increase in the number density of stars.
    \item \textbf{Bit 256, Very bright stars:} very bright stars that produce significant scattered light artifacts were explicitly masked to remove areas with high densities of objects with anomalous colors. These stars are listed in \tabref{famousstars}.
\end{itemize}

\begin{deluxetable}{c c l}
\tablewidth{0pt}
\tabletypesize{\scriptsize}
\tablecaption{ \gold Foreground Region Mask \label{tab:foregrounds}}
\tablehead{
\colhead{Flag Bit} & \colhead{Area ($\deg^2$)}  & \colhead{Description}
}
\startdata
1   &  36.4 & \Gaia bright stars ($G<7$) \\
2   &  19.7 & Yale bright stars  \\
4   &  78.3 & 2MASS bright stars ($4< J < 8$) \\
8  &  158.1 & \Gaia moderately bright stars ($7 < G < 11.5$) \\
16  & 254.6 & 2MASS moderately bright stars ($8< J < 12$) \\
32  & 18.9 & HyperLEDA bright galaxies \\ 
64 &  0.4 & Milky Way satellites (\tabref{satellites}) \\
128 & 102.8 & Large Magellanic Cloud periphery \\
256 & 76.2 & Very bright stars  (\tabref{famousstars})\\
\enddata
\tablecomments{
The masked area from the \gold catalog is calculated using the coverage fraction of the pixels that are removed from the footprint by each mask. The rationale for each mask can be found in \secref{foregrounds}.
}
\end{deluxetable}

The  magnitude-dependent radii for the Gaia, Yale, and 2MASS masks were defined based on a cumulative plot of the ratio of objects with inconsistent \texttt{BDF} and \texttt{AUTO} magnitudes as a function of distance to the bright sources. These ``bad'' objects denoted image artifacts that clustered around the central pixels and radius was defined by inspection so that they are not dominant. Further quality cuts with \flagsgold eliminate additional objects outside these masks. The LMC and very bright star masks were defined on an ad-hoc basis by visual inspection.
\begin{deluxetable}{c c c}
\tablewidth{0pt}
\tabletypesize{\scriptsize}
\tablecaption{\label{tab:satellites} Milky Way globular clusters and satellite galaxy exclusion list. }
\tablehead{
\colhead{Name} & \colhead{$\alpha_{2000}$, $\delta_{2000}$} & \colhead{Radius} \\[-0.5em]
 & (deg, deg) & (deg)
}
\startdata
AM\,1     & $(58.7612, -49.6144)$ & $0.015$ \\
Eridanus  & $(66.1854, -21.1869)$ & $0.015$ \\
Fornax    & $(39.9971, -34.4492)$ & $0.7$ \\
NGC\,0288 & $(13.1979, -26.59)$ & $0.2$ \\
NGC\,1261 & $(48.0637, -55.2169)$ & $0.15$ \\
NGC\,1851 & $(78.5262, -40.0472)$ & $0.2$ \\
NGC\,1904 & $(81.0442, -24.5242)$ & $0.17$ \\
NGC\,7089 & $(323.375,  -0.8167)$ & $0.22$ \\
Reticulum & $(69.0375 , -58.85833)$ & $0.08$ \\
Sculptor  & $(15.03875, -33.7092)$ & $0.7$ \\
Whiting 1 & $(30.7375 , -3.25277)$ & $0.015$ \\
\enddata
\tablecomments{AM1, Eridanus, and Whiting 1 have angular sizes that are smaller than a single \nside=4096 \healpix pixel. Thus, their mask radius is set to the approximate angular size of a pixel.}
\end{deluxetable}

\begin{deluxetable}{c c c}
\tablewidth{0pt}
\tabletypesize{\scriptsize}
\tablecaption{ Very bright stars included in the foreground mask. Third column indicates the masking radius applied for each case. \label{tab:famousstars}}
\tablehead{
\colhead{Name} & \colhead{$\alpha_{2000}$, $\delta_{2000}$} & \colhead{Radius} \\[-0.5em]
 & (deg, deg) & (deg)
}
\startdata
$\alpha$ Phe   & ($6.5708$, $-42.3061$) & $2.0$\\
$\alpha$ Eri   & ($24.4288$, $-57.2367$) & $1.7$\\
$\alpha$ Hyi   & ($29.6925$, $-61.5697$) & $0.5$\\
$\alpha$ Col   & ($84.9121$, $-34.0741$) & $1.0$\\
$\alpha$ Car   & ($95.9879$, $-52.6958$) & $2.5$\\
$\alpha$ Pav   & ($306.41214$, $-56.7350$) & $1.7$\\
$\alpha$ Gru   & ($332.0583$, $-46.9611$) & $1.5$\\
$\beta$ Gru   & ($340.6671$,$-46.8847$) & $2.0$\\
Pi1 Gru & ($335.6829$, $-45.9478$) & $0.5$\\
R Dor & ($69.1900$, $-62.0775$) & $0.5$\\
\enddata
\end{deluxetable}

\begin{figure*}
\includegraphics[width=\textwidth]{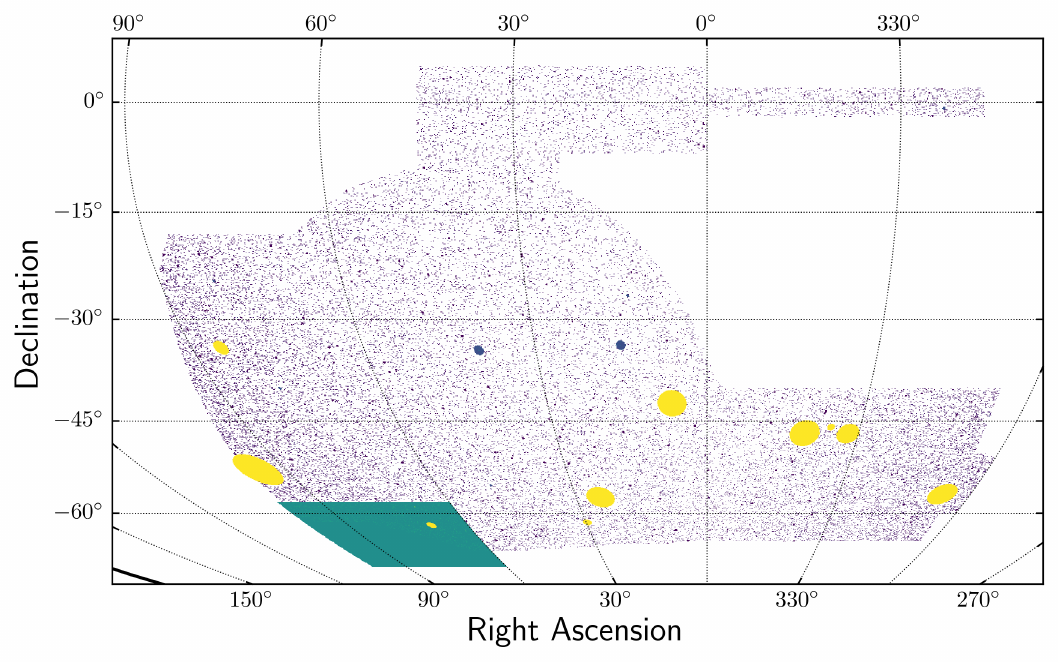}
\caption{\label{fig:foregrounds}
The foreground mask for \gold. The region close to the LMC appears in turquoise, very bright stars appear in yellow \tabref{famousstars}), while the Milky Way satellite galaxies Fornax and Sculptor can be seen in dark blue at ${\rm Dec.} \sim -35\deg$. Small regions around a variety of bright stars and nearby galaxies are masked in purple. See \secref{foregrounds} for details.}
\end{figure*}

\subsection{Survey Properties}
\label{sec:properties}

Survey property maps represent spatially varying distributions of observation characteristics and astrophysical line-of-sight effects that systematically impact the detection and measurement of sources across the footprint, and consequently affect statistical analysis of the large-scale distribution of galaxies \citep[e.g.,][]{2022MNRAS.511.2665R}. 
The \gold survey property maps are distributed in both \healpix and in \healsparse formats.
In both cases, the map content is derived from \mangle polygon masks that encode the full-resolution coadd image geometry.
The full list of survey property maps can be found in \appref{survey_properties}.

\section{Caveats and Known Issues}
\label{sec:issues}

\subsection{Background offset}

\gold photometry is impacted at a low level by both global and local background over-subtraction that is likely attributed to the extended PSF of bright stars and galaxies, coupled with the spatial scale of sky background estimation.
The effect can be recognized through ratios of \sextractor aperture fluxes measured for two different aperture radii averaged over a large number of test stars.
In the case of ideal background modeling and a fixed PSF, this ratio should be independent of both the flux of the stars used for the test and their spatial location within the footprint. 
For \gold, we find that the ratio of aperture fluxes exhibits a flux dependence, with the large-aperture photometry of fainter stars being more impacted by background over-subtraction.
The amplitude of the background offset is correlated with the density of bright stars across the survey footprint.
The largest over-subtraction occurs at an angular separation of $1-2 \amin$ around bright stars and galaxies, corresponding to the $1.1 \amin \times 1.1 \amin$ gridding scale used for background estimation.
Analysis suggests that the extended wings of the PSF are being treated as a background, resulting in over-subtraction relative to the natural sky level.

\subsection{Spurious sources and catastrophic measurement errors}

Removing super-spreader objects (\flagsgold = 8) was found to be potentially problematic for the analyses of galaxies in the central regions of galaxy clusters. Some modifications were introduced to mitigate this problem in the final \gold catalog. However, specific care is advised for the study of galaxies in dense environments, including comparative tests with and without the super-spreader cut. 

\subsection{FITVD Failures}

Three tiles (DES0456-5705, DES0456-5705, DES0424-3249) experienced partial corruption during the main \fitvd run. 
This led to spatially correlated failures of the \fitvd measurements.
Some cosmology papers \citep[e.g.,][]{Y6BAO} used an earlier version of \gold that masked out these three tiles, corresponding to a loss of $1.5 \deg^2$ (0.03\% of the footprint area).
Complete measurements for those three tiles were restored for the final version of \gold released here.

\subsection{Extinction}
\label{sec:extinction}

As described in \secref{reddening}, the $R_b$ coefficients provided by the DES data releases apply a renormalization of $N = 0.78$ to the measured $E(B-V)$ values from SFD98. 
This renormalization was originally suggested by \citet{2010ApJ...725.1175S} and was later used to calculate the $R_b$ values in Table 6 of \citet{2011ApJ...737..103S}. 
However, \citet{2011ApJ...737..103S} suggested that a renormalization of $N = 0.86$ may be more appropriate in low-reddening regions that have $E(B-V) < 0.2$, which is the case for most of the DES footprint. 
Users may easily rescale the $R_b$ extinction values provided by \gold with their preferred renormalization of the SFD98 maps.

\section{Using Y6 Gold}
\label{sec:using}

The \gold data products and user documentation will be released at \drurl \, alongside previous major DES releases. A selection of the most important columns of the catalog is provided in \appref{catalog_columns}.
\gold includes the value-added object catalog together with maps detailed in \secref{maps} in \healpix and \healsparse formats.

General usage recommendations are listed below. 

\begin{itemize}
   \item Use \textbf{$\flagsfootprint = 1$} to select objects located within the standard \gold footprint, as described in \secref{footprint}.
   \item Regions with astrophysical foregrounds identified in \secref{foregrounds} can present various problems in terms of photometry, spurious detections, obscuration, etc. The \textbf{$\flagsforeground = 0$} selection is generally recommended for extragalactic studies.
   \item As explained in \secref{flagsgold}, \flagsgold facilitates the selection of good quality objects by summarizing various flags and signatures of poor reconstructions in a single bitmask. A \textbf{$\flagsgold = 0$} selection will suffice for most applications.
   \item All photometry measurements include atmospheric and instrumental calibration derived from FGCM (i.e., top-of-the-atmosphere photometry; \secref{calibration}). 
   \textbf{By default, reported fluxes/magnitudes are NOT corrected for interstellar dust extinction.}
   Final top-of-the-Galaxy photometry can be obtained by applying an aperture correction and fiducial de-reddening (\secref{photometric_corrections}); only the measurements with the \textbf{\texttt{\_CORRECTED}} suffix take into account these two adjustments.
   \item \textbf{The \texttt{EXT\_MASH} star/galaxy separator is expected to be appropriate for most scientific applications}. This classifier is based on morphological quantities, as described in \secref{classification} and \appref{extclass}. The method employs \var{EXT\_FITVD} as the main classifier for an object, but reverts to \var{EXT\_WAVG} or \var{EXT\_COADD} measurements as necessary. For cosmology analyses, the selection \textbf{$\var{EXT\_MASH} = 4$} is a recommended starting point, since it shows low stellar contamination up to the magnitude limit, with a decrease in galaxy selection efficiency only beyond $i > 22.5$\,mag. The outputs of the \code{XGBoost} classifier (\var{EXT\_XGB} and \var{XGB\_PRED}) have been found to outperform \var{EXT\_MASH}, but have received much less validation in the context of cosmological analyses. 
\end{itemize}

The \gold data used for most Y6 cosmology analyses corresponds to DES internal version 2.2. 

\section{Summary and Outlook}
\label{sec:conclusions}

The \gold data products presented here, together with weak lensing shear (Yamamoto \& Becker et al., in prep.) and Deep Field (Gruendl et al., in prep.) catalogs, form the foundation of legacy static-sky cosmology from the full observational data set of DES.
Components of the final \gold release, summarized in \tabref{summary}, include:

\begin{itemize}
    \item A catalog of \ngold high-quality objects covering ${\sim}5000\,\deg^2$ of the southern Galactic cap to a depth of $i_{AB}\sim  \maglimsofiapp$ at $\SNR \sim 10$ for extended objects with measurements in the $grizY$ bands derived from the DES Wide-Field Survey data released in DES DR2 \citep{DESDR2}.
    \item Flux measurements for PSF, \bdf, and \gap models derived from simultaneous fits to multi-epoch, multi-band photometry to enable more robust determination of colors and morphology.
    \item Per-object aperture corrections and interstellar extinction estimates to take full advantage of top-of-the-atmosphere photometric uniformity of $<2$~mmag.
    \item Improved morphological object classification schemes based on both conventional and machine-learning approaches.
    \item Photometric redshifts derived with the DNF estimator \citep{dnf} .
    \item An expanded set of per-object flags to select reliable object samples.
    \item Foreground mask to select recommended regions for extragalactic studies.
    \item High-resolution footprint and survey property maps representing the observational coverage and properties of the DES data set.
\end{itemize}

\noindent These curated and validated data products will enable some of the tightest constraints on the standard cosmological model to date, and are well suited for detailed statistical analyses of extragalactic populations and the Milky Way stellar halo. 
Data products and documentation are publicly available at \drurl.

A new generation of wide-area imaging surveys will soon advance our understanding of new physics implied by cosmological observations and theory.
Ground-based surveys including the Vera C.\ Rubin Observatory's LSST  will catalog $>10^{10}$ galaxies and $>10^5$ Type Ia supernovae \citep[][]{Ivezic:2019}.
The \emph{Euclid} \citep{2024arXiv240513491E} and \emph{Nancy Grace Roman} \citep{roman} observatories will use high-resolution space-based imaging to cover complementary spatial regions, depth ranges, and wavelengths.
Meeting the statistical grasp of these new projects to make accurate cosmological inferences will require even more stringent control of systematic effects related to the detectors, atmosphere, and survey observations.
DES has been an important development and testing ground for pixel-level processing, calibration, and measurement algorithms, several of which are now being incorporated into the LSST Science Pipelines \citep{bosch2018,bosch2019} including methods for representing survey geometry and metadata (\code{healsparse}), PSF modeling (\PIFF), photometric calibration (FGCM), astrometric calibration with simultaneous solution across bands and coadd input images, survey-scale synthetic source injection (\code{Balrog}), cell-based coaddition, weak lensing shear measurement (\code{metadetection}, \code{BFD}), and usage of deep field processing for accurate shape and color references for the wide survey.
Compelling science questions, new observational capabilities, and continuously improving methods for data management and analysis promise an exciting future for wide-area imaging surveys for years to come.

\acknowledgments

We thank Lynne Jones and Jean-Charles Cuillandre for help with the LSST and Euclid footprints. 

Contribution statement: KB led the \gold development and contributed to manuscript writing; ISN contributed to catalog and mask creation and manuscript writing; ADW developed the XGB star/galaxy classifier and contributed to manuscript writing; BY \& RG led many technical aspects of the DES Data Management production and contributed to manuscript writing; ES designed the \fitvd and \ngmix algorithms and contributed to the manuscript. 
Authors ER, JdV, MA, DA, MRB, GB, ACR, JG, MG, WGH, MJ, TJ, RK, AM, JO, AP, MRM, DSC, MT, LTSC, DLT, NW, and MY contributed to various aspects of assembling and validating \gold including the creation of specific data products, scientific validation, manuscript writing, manuscript review, data product documentation, and data release, as well as coordinating the processes mentioned above.
The remaining authors have made contributions to this paper that include, but are not limited to, the construction of DECam and other aspects of collecting the data; data processing and calibration; developing broadly used methods, codes, and simulations; running the pipelines and validation tests; and promoting the science analysis.
WGH \& RK served as internal reviewers for the manuscript.

Funding for the DES Projects has been provided by the U.S. Department of Energy, the U.S. National Science Foundation, the Ministry of Science and Education of Spain, the Science and Technology Facilities Council of the United Kingdom, the Higher Education Funding Council for England, the National Center for Supercomputing 
Applications at the University of Illinois at Urbana-Champaign, the Kavli Institute of Cosmological Physics at the University of Chicago, the Center for Cosmology and Astro-Particle Physics at the Ohio State University,
the Mitchell Institute for Fundamental Physics and Astronomy at Texas A\&M University, Financiadora de Estudos e Projetos, Funda{\c c}{\~a}o Carlos Chagas Filho de Amparo {\`a} Pesquisa do Estado do Rio de Janeiro, Conselho Nacional de Desenvolvimento Cient{\'i}fico e Tecnol{\'o}gico and the Minist{\'e}rio da Ci{\^e}ncia, Tecnologia e Inova{\c c}{\~a}o, the Deutsche Forschungsgemeinschaft and the Collaborating Institutions in the Dark Energy Survey. 

The Collaborating Institutions are Argonne National Laboratory, the University of California at Santa Cruz, the University of Cambridge, Centro de Investigaciones Energ{\'e}ticas, Medioambientales y Tecnol{\'o}gicas-Madrid, the University of Chicago, University College London, the DES-Brazil Consortium, the University of Edinburgh, the Eidgen{\"o}ssische Technische Hochschule (ETH) Z{\"u}rich, Fermi National Accelerator Laboratory, the University of Illinois at Urbana-Champaign, the Institut de Ci{\`e}ncies de l'Espai (IEEC/CSIC), 
the Institut de F{\'i}sica d'Altes Energies, Lawrence Berkeley National Laboratory, the Ludwig-Maximilians Universit{\"a}t M{\"u}nchen and the associated Excellence Cluster Universe, the University of Michigan, NSF NOIRLab, the University of Nottingham, The Ohio State University, the University of Pennsylvania, the University of Portsmouth, SLAC National Accelerator Laboratory, Stanford University, the University of Sussex, Texas A\&M University, and the OzDES Membership Consortium.

Based in part on observations at NSF Cerro Tololo Inter-American Observatory at NSF NOIRLab (NOIRLab Prop. ID 2012B-0001; PI: J. Frieman), which is managed by the Association of Universities for Research in Astronomy (AURA) under a cooperative agreement with the National Science Foundation.

The DES data management system is supported by the National Science Foundation under Grant Numbers AST-1138766 and AST-1536171. The DES participants from Spanish institutions are partially supported by MICINN under grants PID2021-123012, PID2021-128989 PID2022-141079, SEV-2016-0588, CEX2020-001058-M and CEX2020-001007-S, some of which include ERDF funds from the European Union. IFAE is partially funded by the CERCA program of the Generalitat de Catalunya.

We  acknowledge support from the Brazilian Instituto Nacional de Ci\^enciae Tecnologia (INCT) do e-Universo (CNPq grant 465376/2014-2).






This work has made use of data from the European Space Agency (ESA) mission {\it Gaia} (\url{https://www.cosmos.esa.int/gaia}), processed by the {\it Gaia} Data Processing and Analysis Consortium (DPAC,
\url{https://www.cosmos.esa.int/web/gaia/dpac/consortium}). Funding for the DPAC has been provided by national institutions, in particular the institutions
participating in the {\it Gaia} Multilateral Agreement.

This paper makes use of observations obtained as part of the VISTA Hemisphere Survey, ESO Program, 179.A-2010 (PI: McMahon).

The Hyper Suprime-Cam (HSC) collaboration includes the astronomical communities of Japan and Taiwan, and Princeton University. The HSC instrumentation and software were developed by the National Astronomical Observatory of Japan (NAOJ), the Kavli Institute for the Physics and Mathematics of the Universe (Kavli IPMU), the University of Tokyo, the High Energy Accelerator Research Organization (KEK), the Academia Sinica Institute for Astronomy and Astrophysics in Taiwan (ASIAA), and Princeton University. Funding was contributed by the FIRST program from the Japanese Cabinet Office, the Ministry of Education, Culture, Sports, Science and Technology (MEXT), the Japan Society for the Promotion of Science (JSPS), Japan Science and Technology Agency (JST), the Toray Science Foundation, NAOJ, Kavli IPMU, KEK, ASIAA, and Princeton University. 

This paper is based on data collected at the Subaru Telescope and retrieved from the HSC data archive system, which is operated by the Subaru Telescope and Astronomy Data Center (ADC) at NAOJ. Data analysis was in part carried out with the cooperation of Center for Computational Astrophysics (CfCA), NAOJ. We are honored and grateful for the opportunity of observing the Universe from Maunakea, which has cultural, historical and natural significance in Hawaii. 

This paper makes use of software developed for Vera C. Rubin Observatory. We thank the Rubin Observatory for making their code available as free software at http://pipelines.lsst.io/.



This manuscript has been authored by Fermi Research Alliance, LLC under Contract No. DE-AC02-07CH11359 with the U.S. Department of Energy, Office of Science, Office of High Energy Physics.

\vspace{5mm}
\facility{Blanco (DECam)} 

\software{
\code{astropy} \citep{Astropy:2013}, 
\code{decasu},\footnote{\url{https://github.com/erykoff/decasu}}
\code{DNF} \citep{dnf},\footnote{\url{https://github.com/ltoribiosc/DNF_photoz}}
\code{easyaccess}\citep{easyaccess}, 
\code{fitsio},\footnote{\url{https://github.com/esheldon/fitsio}}
\code{fpack} \citep{fpack},
\healpix \citep{HealpixSoft},\footnote{\url{http://healpix.sourceforge.net}}
\code{healpy} \citep{Zonca2019},\footnote{\url{https://github.com/healpy/healpy}}
\code{healsparse},\footnote{\url{https://healsparse.readthedocs.io/en/latest/}}
\mangle \citep{2004MNRAS.349..115H,2008MNRAS.387.1391S}, 
\code{matplotlib} \citep{2007CSE.....9...90H}, 
\code{numpy} \citep{numpy:2011}, 
\PSFEx \citep{psfex}, 
\PIFF \citep{piff},
\scamp \citep{2006ASPC..351..112B}, 
\code{scipy} \citep{scipy:2001}, 
\code{skyproj},\footnote{\url{https://github.com/LSSTDESC/skyproj}}
\SExtractor \citep{sextractor}, 
\swarp \citep{2002ASPC..281..228B, 2010ascl.soft10068B}, 
\code{TOPCAT} \citep{2005ASPC..347...29T},
\code{XGBoost} \citep{Chen:2016}.
}

\appendix
\numberwithin{figure}{section}
\numberwithin{table}{section}

\section{Object Classification}
\label{app:classification}

The DES \gold morphological objects classification scheme follows from similar schemes developed for DES Y3 Gold \citep[Section 6.1 and Appendix B in][]{Y3Gold} and DES DR2 \citep[Section 4.7 in][]{DESDR2}. This approach defines independent object classes based on the multi-epoch \fitvd measurements, the weighted average of the \code{SourceExtractor} measurements on the individual images, and the \code{SourceExtractor} measurements on the coadded images. These independent classifications are then combined hierarchically to provide a single classification for every object in DES \gold. In addition, DES \gold includes a gradient boosted decision tree model that incorporates morphological information in an automated classification procedure. The continuously valued output of the \xgboost model is divided into discrete object classes that roughly match the completeness of the conventional classifier.  We discuss each of these classification approaches in more detail below.

\subsection{Conventional Classifiers}
\label{app:extclass}

Independent object classifications are derived from the multi-epoch \fitvd measurements (\code{EXT\_FITVD}), the weighted average of \code{SourceExtractor} measurements on the individual images (\code{EXT\_WAVG}), and the \code{SourceExtractor} measurements on the coadded images (\code{EXT\_COADD}). Each classifier assigns an integer value from 0 to 4, with 0 being high-confidence stars/QSOs and 4 being high-confidence galaxies. When the class cannot be computed based on the specific measurement technique, a default value of $-9$ is assigned. These independent classifications are combined heirarchically to provide a classification for every object in DES \gold (\code{EXT\_MASH}). 

\begin{figure*}
    \centering
    \includegraphics[width=0.49\textwidth]{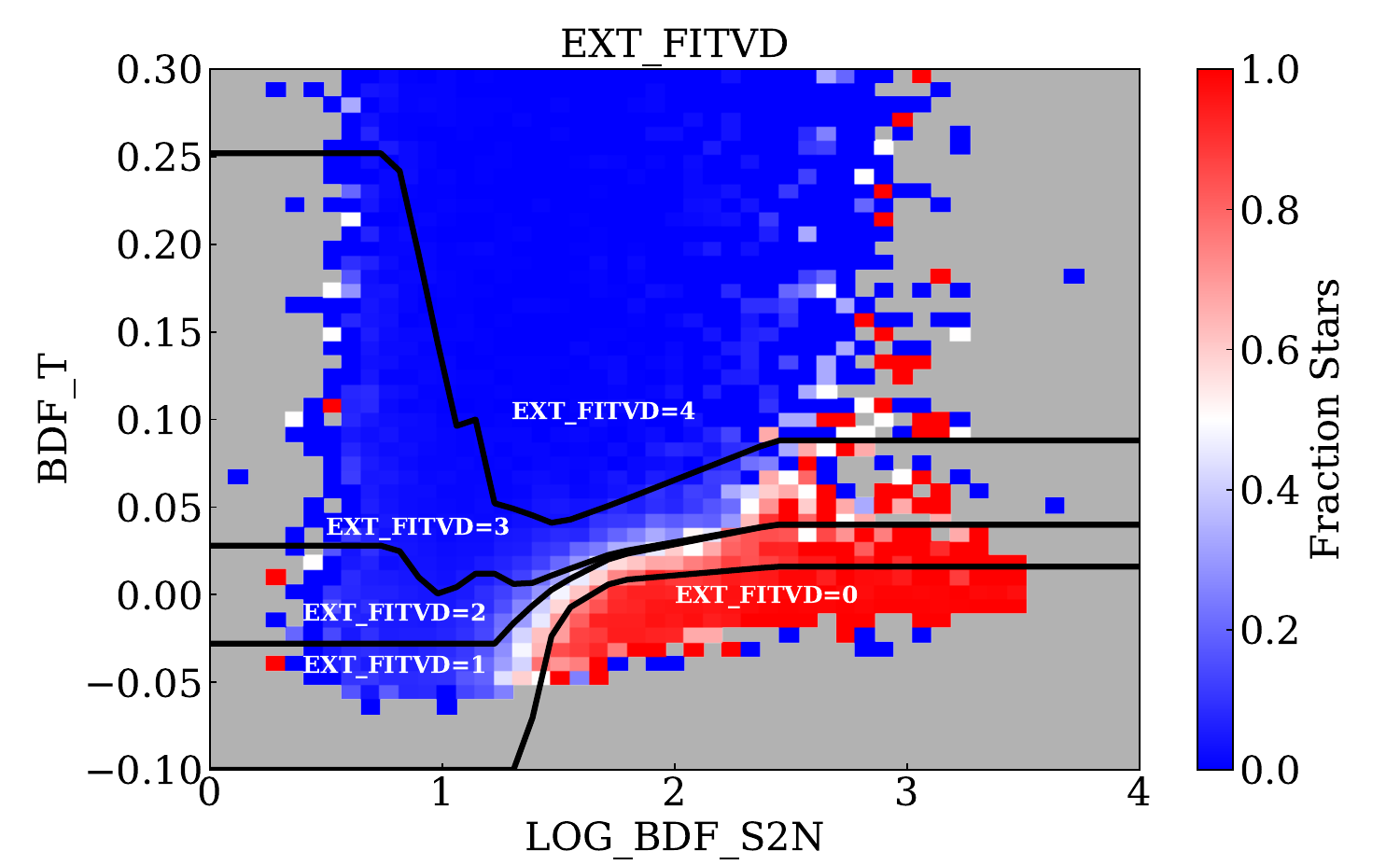}
    \includegraphics[width=0.49\textwidth]{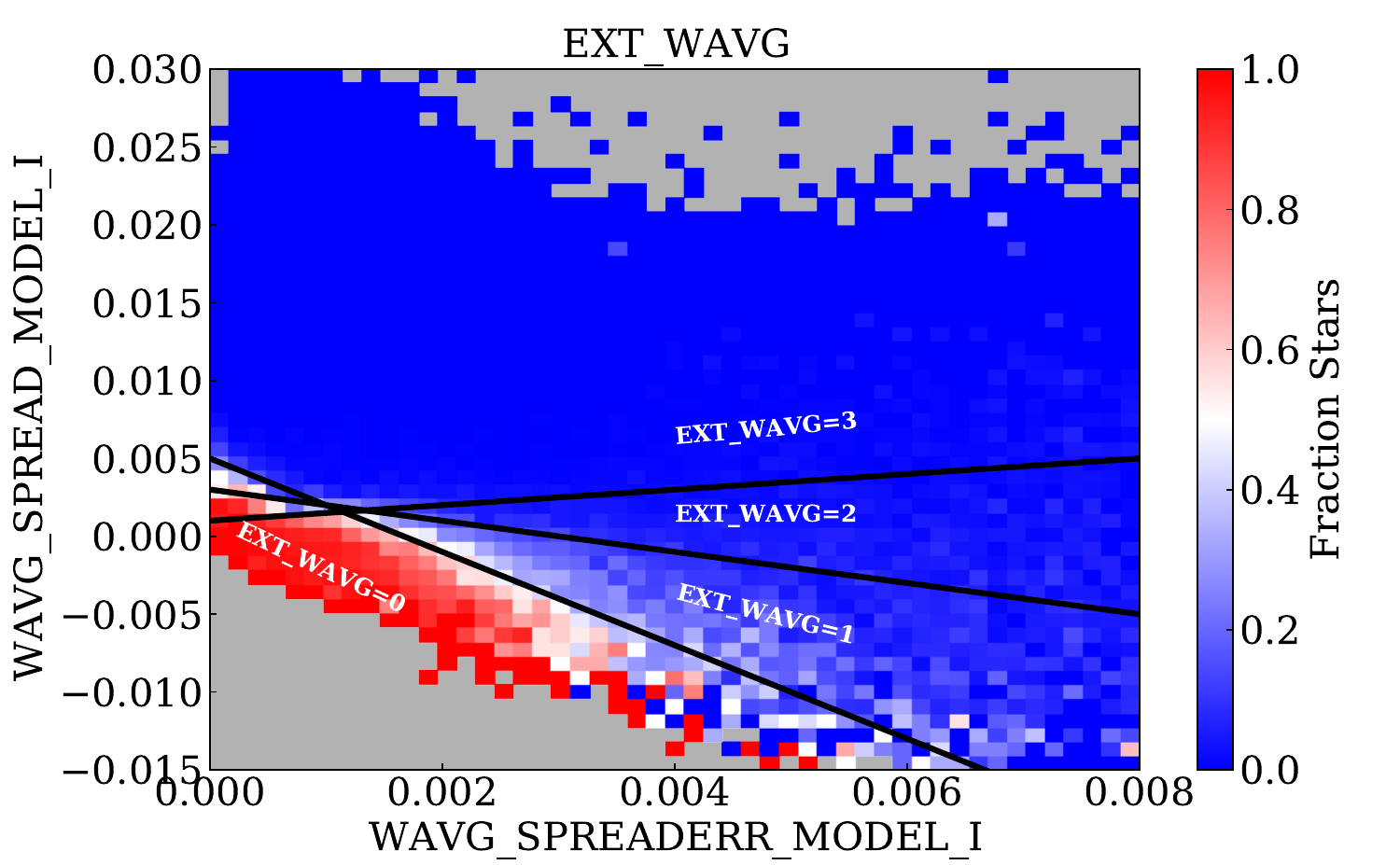} 
    \includegraphics[width=0.49\textwidth]{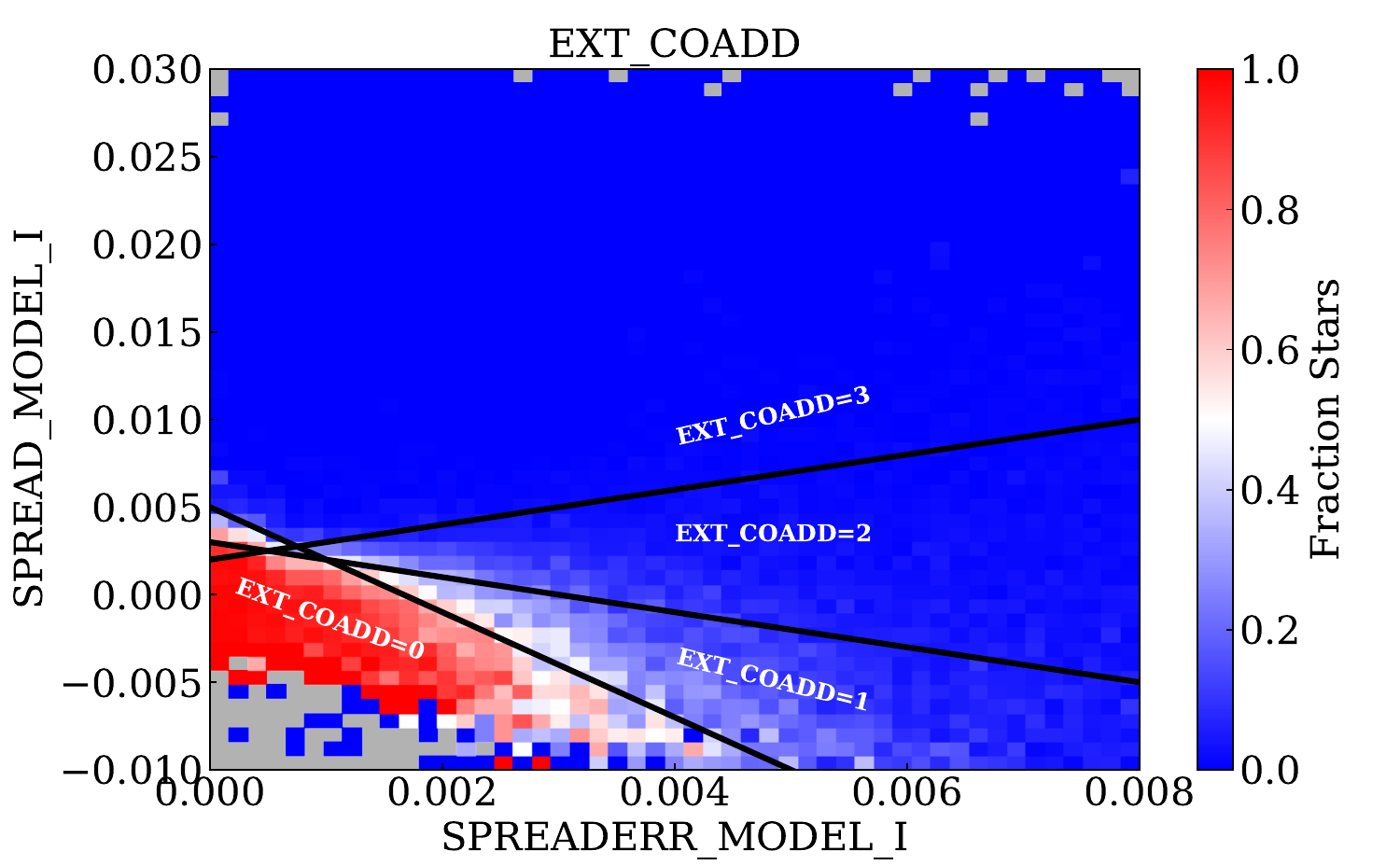}
    \caption{Conventional star/galaxy clases are defined in pairs of morphological parameters from the \fitvd (left), \code{WAVG} (middle), or \code{COADD} (right) measurements. Each panel shows the two-parameter space where the extended classes are defined. The background color indicates the fraction of objects classified as stars using a morphological selection from HSC PDR2 and infrared data from CLAUDS in the XMM-LSS field \citep{Desprez:2023}. \label{fig:extclass}}
\end{figure*}

The \fitvd extended classifier (\code{EXT\_FITVD}) is based on a series of cuts in the space of measured size (\code{BDF\_T}) vs.\ signal-to-noise ($\log_{10}(\code{BDF\_S2N})$) (\figref{extclass}). In this space, a set of precision-recall curves (i.e., completeness versus purity) were created by varying the \code{BDF\_T} threshold to separate stars and galaxies. The cuts in this space were based on the threshold value that maximized the Matthews Correlation Coefficient giving equal weights to stars and galaxies, and alternatives that gave higher weights to the purity of target populations.
The threshold for each class is expressed as a linear interpolation function, $f$, that returns the threshold on \code{BDF\_T} ($y$) as a function of $\log_{10}(\code{BDF\_S2N})$ ($x$). The values for these interpolation functions are shown in \tabref{extfitvd}. The integer value of the \code{EXT\_FITVD} classifier is then defined as the sum of the thresholds that the object exceeds.
\begin{equation}
    \code{EXT\_FITVD} = \sum_i^4 \big \{ \code{BDF\_T} > f_i(\log_{10}(\code{BDF\_S2N}) ) \big \}.
\end{equation}

\begin{deluxetable}{ccccc}
\tablecaption{ Interpolation nodes for \code{EXT\_FITVD}. \label{tab:extfitvd}}

\tablewidth{6in}
\tablehead{\colhead{$x$} & \colhead{$~y_1~$} & \colhead{$~y_2~$} & \colhead{$~y_3~$} & \colhead{$~y_4~$}}
\startdata
   -3.0       & $-0.1$   & $-0.028$ &  0.028   & 0.252 \\
   0.79891862 & $-0.1$   & $-0.028$ &  0.028   & 0.252 \\
   0.90845217 & $-0.1$   & $-0.028$ &  0.008   & 0.188 \\
   0.98558583 & $-0.1$   & $-0.028$ &  0.      & 0.14  \\
   1.05791208 & $-0.1$   & $-0.028$ &  0.004   & 0.096 \\
   1.13603715 & $-0.1$   & $-0.028$ &  0.012   & 0.104 \\
   1.22479487 & $-0.1$   & $-0.028$ &  0.012   & 0.052 \\
   1.33572223 & $-0.1$   & $-0.012$ &  0.004   & 0.048 \\
   1.48983602 & $-0.012$ & 0.005    &  0.012   & 0.04  \\ 
   1.74124395 & 0.008    & 0.022    &  0.024   & 0.052 \\
   2.43187589 & 0.016    & 0.04     &  0.04    & 0.088 \\
   6.0        & 0.016    & 0.04     &  0.04    & 0.088 \\
\enddata
\end{deluxetable}

The weighted average extended classifier (\code{EXT\_WAVG}) is built from the weighted average of the \code{SourceExtractor} $i$-band \code{SPREAD\_MODEL} and \code{SPREADERR\_MODEL} measurements from the individual exposures. This classifier makes use of the most accurate PSF for each individual exposure, but is limited to the depth of a single exposure.
\begin{align}
    \begin{split}
    \code{EXT\_WAVG}  &= ((\code{WAVG\_SPREAD\_MODEL\_I} + 3.0*\code{WAVG\_SPREADERR\_MODEL\_I}) > 0.005) \\
              &+ ((\code{WAVG\_SPREAD\_MODEL\_I} + 1.0*\code{WAVG\_SPREADERR\_MODEL\_I}) > 0.003) \\
              &+ ((\code{WAVG\_SPREAD\_MODEL\_I} - 0.5*\code{WAVG\_SPREADERR\_MODEL\_I}) > 0.001)
    \label{eqn:ext_wavg}
    \end{split}
\end{align}

The coadd extended classifier (\code{EXT\_COADD}) is built from the \code{SourceExtractor} measurements of \code{SPREAD\_MODEL} and \code{SPREADERR\_MODEL} on the $i$-band coadd images. This is the most complete classifier (returning a value for nearly every object), but it suffers from the limitations of the coadded image PSF that is subject to discontinuities and sharp variations in depth. For this reason, it is given the lowest priority.

\begin{align}
    \begin{split}
    \code{EXT\_COADD}  &= ((\code{SPREAD\_MODEL\_I} + 3.0*\code{SPREADERR\_MODEL\_I}) > 0.005) \\
              &+ ((\code{SPREAD\_MODEL\_I} + 1.0*\code{SPREADERR\_MODEL\_I}) > 0.003) \\
              &+ ((\code{SPREAD\_MODEL\_I} - 1.0*\code{SPREADERR\_MODEL\_I}) > 0.002)
    \label{eqn:ext_coadd}
    \end{split}
\end{align}

The combined extended classifier, \code{EXT\_MASH}, is assembled from the combination of \code{EXT\_FITVD}, \code{EXT\_WAVG}, and \code{EXT\_COADD} classifications.

\begin{equation}
    \code{EXT\_MASH}  = \left\{\begin{array}{ll}
        \code{EXT\_FITVD}, & \text{~if } \code{EXT\_FITVD} > -9 \\
        \code{EXT\_WAVG},  & \text{~elif } \code{EXT\_WAVG} > -9 \\
        \code{EXT\_COADD}, & \text{~elif } \code{EXT\_COADD} > -9 \\
        -9,                & \text{~otherwise}
        \end{array}\right\}
        \label{eqn:ext_mash}
\end{equation}

\subsection{\xgboost Classifier}
\label{app:xgboost}

\begin{deluxetable}{cl}
\tablecaption{Input parameters to the \xgboost star/galaxy classifier. \label{tab:xgboost}}
\tablehead{\colhead{Variable Name} & \colhead{Description}}
\startdata
   \code{CONC} & Concentration parameter from DESCONC \\
   \code{BDF\_T} & Multi-epoch buldge + disk fit size parameter \\
   $\log_{10}$(\code{BDF\_S2N}) & Logarithm of the multi-epoch bulge+disk fit signal-to-noise \\
   \code{BDF\_T\_ERR} & Uncertainty on the multi-epoch buldge+disk fit size parameter \\
   \code{WAVG\_SPREAD\_MODEL\_I} & weighted average \code{SourceExtractor} \code{SPREAD\_MODEL} in $i$-band \\
   \code{WAVG\_SPREADERR\_MODEL\_I} & weighted average \code{SourceExtractor} \code{SPREADERR\_MODEL} in $i$-band \\
\enddata
\end{deluxetable} 

Machine learning provides another well-tested approach to problems of star/galaxy classification \citep[e.g.,][]{2019MNRAS.483..529C}. In the context of DES Y1, a wide variety of ML models were explored for star/galaxy classification \citep{Sevilla-Noarbe:2018}. Here, we apply the popular gradient-boosted decision tree algorithm, \xgboost \citep{Chen:2016}, to perform a classification of stars and galaxies in DES \gold. Our training sample is assembled from two high-purity samples covering the bright and faint ends of the DES catalog. 
At the bright end, we use a combination of Gaia EDR3 morphology \citep{gaiaedr3} and SDSS DR17 spectral classifications \citep{sdssdr17}. 
At the faint end, we used the combination of HSC-SSP PDR2 morphology and CLAUDS infrared colors assembled in the XMM-LSS field by \cite{Desprez:2023}. 
In assembling these ``truth'' labels for training the \xgboost classifier, we were specifically focused on the purity of our samples rather than their completeness.

The \xgboost model was trained on a set of six parameters listed in \tabref{xgboost}. 
During training, the input data set was augmented with a small ($5\%$) sample where one or more of the input parameters were explicitly set as missing. The \xgboost algorithm can deal with missing values through leaf trifurcation and was thus trained to be robust against one or more missing measurements in the real data.
Optimization of the \xgboost hyperparameters was explored using scikit-learn \code{RandomSearchCV}.
The specific scientific focus when designing the \xgboost classifier was on maximizing the completeness and purity of the {\it stellar} sample at faint magnitudes; however, it was also found to deliver excellent performance for galaxies. 

The output of the \xgboost classifier is a continuous valued variable (\code{XGB\_PRED}). Values of $\code{XGB\_PRED} \sim 0$ indicate extended (galaxy-like) objects,  while values of $\code{XGB\_PRED} \sim 1$ indicate point-like (stellar) objects. Following the convention of the classical cut-based classifiers (\appref{extclass}), the sample of objects was divided into discrete classes enumerated by integer values of $\{-9, 0, 1, 2, 3, 4\}$ by placing cuts on the \code{XGB\_PRED} output. 
The placement of these cuts was designed to return approximately the same number of objects as the equivalent \code{EXT\_MASH} class when applied to the full DES \gold object catalog. Again, a value of $\code{EXT\_XGB} = -9$ indicates no data. In general, the \xgboost-based classifier is found to outperform the conventional classifiers (i.e., giving higher efficiency at fixed contamination, or vice versa, lower contamination at fixed efficiency). However, the \xgboost classificaiton has not been implemented on simulation--injection--recovery tests due to the fact that the \code{CONC} parameter was not measured for the simulated object samples.

\begin{align}
    \begin{split}
    \code{EXT\_XGB}  &= (\code{XGB\_PRED} < 0.865) \\
                     &+ (\code{XGB\_PRED} < 0.110) \\
                     &+ (\code{XGB\_PRED} < 0.045) \\
                     &+ (\code{XGB\_PRED} < 0.015)
    \label{eqn:ext_xgb}
    \end{split}
\end{align}

\subsection{Classifier performance}

\tabref{extclass_performance} summarizes the integrated performance of the conventional and \xgboost star/galaxy classifiers as a function of magnitude and object class.
We provide these performance metrics for two different samples of objects: (1) a relatively bright sample ($17.5 < \var{MAG\_AUTO\_I} < 22.5$) that is intended as a proxy for the galaxy samples used for large-scale structure cosmology analyses, and (2) a more expansive sample similar to what might be considered for more general astronomical analyses ($16.5 < \var{MAG\_AUTO\_I} < 23.5$).
A direct comparison between the conventional \code{EXT\_MASH} and \xgboost classifiers is not possible from this table since the two algorithms classes are not matched on efficiency or contamination for this specific sample of objects. 
However, studies have found that the \xgboost classifier outperforms the \var{EXT\_MASH} classifier when cuts on \var{XGB\_PRED} are set to match the \var{EXT\_MASH} classes on either efficiency (e.g., lower contamination at fixed efficiency) or contamination (e.g., higher efficiency at fixed contamination).

\begin{deluxetable}{l c c c c c c}
\tablewidth{0pt}
\tabletypesize{\tablesize}
\tablecaption{Performance of the morphological star/galaxy separation.
\label{tab:extclass_performance}}
\tablehead{
\colhead{Selection} & \hspace{1cm} & \multicolumn{2}{c}{$17.5 \leq \var{MAG\_AUTO\_I} \leq 22.5$} & \hspace{1cm} & \multicolumn{2}{c}{$16.5 \leq \var{MAG\_AUTO\_I} \leq 23.5$} \\
 & & \colhead{Efficiency} & \colhead{Contamination} & \hspace{1cm} & \colhead{Efficiency} & \colhead{Contamination}}
\startdata
& & \multicolumn{5}{c}{Galaxy Selection} \\
\hline
$2 \leq \var{EXT\_MASH} \leq 4$ & & 99.6\% & 1.3\% & & 99.2\% & 1.8\% \\
$3 \leq \var{EXT\_MASH} \leq 4$ & & 99.6\% & 1.3\% & & 98.8\% & 1.6\% \\
$\var{EXT\_MASH} = 4$           & & 98.6\% & 0.8\% & & 96.3\% & 1.0\% \\
\hline
$2 \leq \var{EXT\_XGB} \leq 4$  & & 99.0\% & 0.5\% & & 97.7\% & 1.0\% \\
$3 \leq \var{EXT\_XGB} \leq 4$  & & 98.3\% & 0.4\% & & 96.2\% & 0.8\% \\
$\var{EXT\_XGB} = 4$            & & 96.7\% & 0.3\% & & 92.5\% & 0.5\% \\
\hline
$\var{XGB\_PRED} \leq 0.65$        & & 99.6\% & 1.1\% & & 99.5\% & 2.1\% \\
$\var{XGB\_PRED} \leq 0.50$        & & 99.6\% & 0.9\% & & 99.3\% & 1.7\% \\
$\var{XGB\_PRED} \leq 0.058$       & & 98.6\% & 0.4\% & & 96.7\% & 0.8\% \\
\hline
& & \multicolumn{5}{c}{Stellar Selection} \\
\hline
$0 = \var{EXT\_MASH}$           & & 89.2\% & 0.7\% & & 80.0\% & 2.0\% \\
$0 \leq \var{EXT\_MASH} \leq 1$ & & 94.6\% & 1.5\% & & 88.9\% & 5.2\% \\
$0 \leq \var{EXT\_MASH} \leq 2$ & & 94.7\% & 1.6\% & & 90.2\% & 7.1\% \\
\hline
$0 = \var{EXT\_XGB}$            & & 92.1\% & 1.0\% & & 79.3\% & 1.5\% \\
$0 \leq \var{EXT\_XGB} \leq 1$  & & 98.0\% & 4.0\% & & 94.3\% & 12.5\% \\
$0 \leq \var{EXT\_XGB} \leq 2$  & & 98.5\% & 6.4\% & & 95.6\% & 19.2\% \\
\hline
$\var{XGB\_PRED} > 0.906$       & & 89.2\% & 0.9\% & & 74.8\% & 1.1\% \\
$\var{XGB\_PRED} > 0.76$        & & 94.6\% & 1.2\% & & 84.2\% & 2.2\% \\
$\var{XGB\_PRED} > 0.75$        & & 94.7\% & 1.3\% & & 84.6\% & 2.2\% \\
\hline
\enddata
\tablecomments{The \var{EXT\_XGB} classes were assigned to give a similar number of objects per class as \var{EXT\_MASH}. The \var{XGB\_PRED} cuts were tuned to give the same completeness as \var{EXT\_MASH} on the specific evaluation data set used.}
\end{deluxetable}

\begin{figure*}[t]
    \centering
    \includegraphics[width=0.55\textwidth]{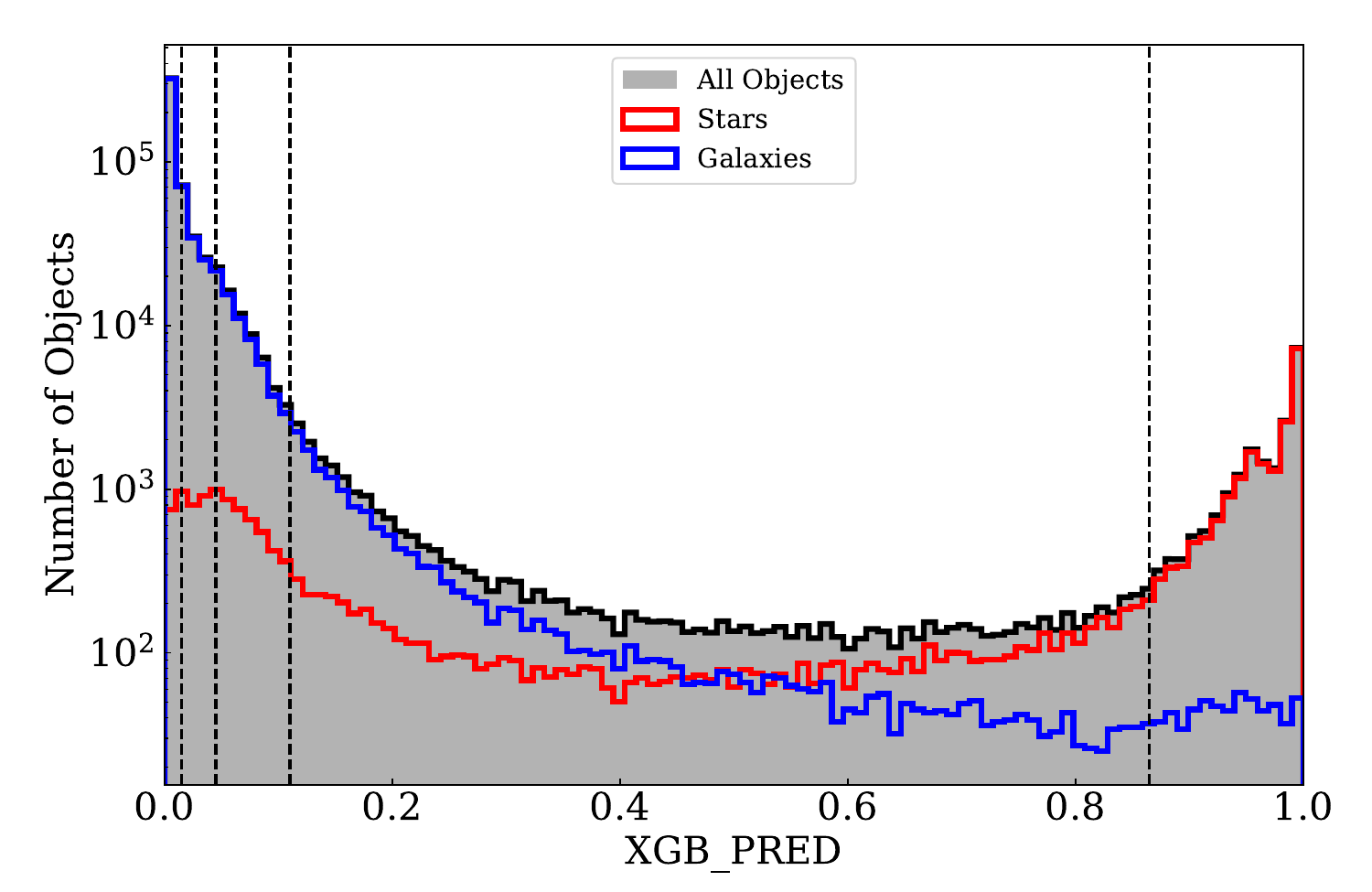}
    \caption{Performance of the \xgboost star/galaxy classifier output. The gray histogram represents the \code{XGB\_PRED} value for all objects in a hold out test sample. The red histogram shows the distribution of objects with the truth label of stars, while the blue histogram shows distribution of objects with a truth label of galaxies. The dashed black lines show the thresholds used to define the \code{EXT\_XGB} classes (4 to 0 from left to right). \label{fig:xgboost}}
\end{figure*}

\section{Main catalog columns}
\label{app:catalog_columns}

\begin{deluxetable}{c c l}[h]
\tablewidth{0pt}
\tabletypesize{\tablesize}
\tablecaption{ Selected \gold catalog columns. \label{tab:catalog_columns}}
\tablehead{
\colhead{\gold catalog column family}  & \colhead{Units} & \colhead{Description} 
}
\startdata
\hline
COADD\_OBJECT\_ID & & Unique identifier for a Y6 coadd object\\
\hline
TILENAME & & Coadd tile to which the object belongs to. See \citet{desdm}.\\ 
\hline
RA, DEC, GLAT, GLON & Degrees & Equatorial and Galactic coordinates\\
\hline
ALPHAWIN\_J2000, DELTAWIN\_J2000 & Degrees & \makecell[cl]{Equatorial coordinates using a Gaussian-windowed \\ measurement (for precise astrometry)}\\
\hline
(BDF/GAP)\_(MAG/FLUX)\_(GRIZY) & \makecell[cc]{Magnitudes\\Counts per s} & \makecell[cl]{Photometry as measured by the \fitvd algorithm, \\ both for Bulge and Disk model or a Gaussian aperture fit}\\
\hline
PSF\_(MAG/FLUX)\_APER8\_(GRIZY) & \makecell[cc]{Magnitudes\\Counts per s} & \makecell[cl]{PSF photometry as measured by the \fitvd algorithm, \\ in APER8 system.} \\
\hline
(BDF/GAP)\_(MAG/FLUX)\_ERR\_(GRIZY)& \makecell[cc]{Magnitudes\\Counts per s} & Estimated error for the BDF/GAP\_(MAG/FLUX)\\
\hline
BDF\_FLUX\_COV\_(1-5)\_(1-5) & \makecell[cc]{Counts per s} &  Elements of the $5 \times 5$ flux covariance matrix for the BDF fit. \\ 
\hline
PSF\_(MAG/FLUX)\_ERR\_APER8\_(GRIZY) & \makecell[cc]{Magnitudes\\Counts per s} & Estimated error to PSF\_(MAG/FLUX)\_APER8\\
\hline
(BDF/GAP)\_(MAG/FLUX)\_(GRIZY)\_CORRECTED & \makecell[cc]{Magnitudes\\Counts per s} &  \makecell[cl]{FLUX corrected for interstellar extinction (i.e. de-reddened; top of Galaxy) \\ and PSF aperture ratio (APER8 system)}\\ 
\hline
PSF\_MAG\_APER8\_(GRIZY)\_CORRECTED & Magnitudes & \makecell[cl] {Magnitude measured by \fitvd PSF model corrected for interstellar extinction \\ (de-reddened top of Galaxy) and PSF aperture ratio (APER8 system).\\ Recommended for point-source studies.} \\
\hline
A\_FIDUCIAL\_(GRIZY) & Magnitudes & \makecell[cl]{SED-independent interstellar extinction based on the $E(B-V)$\\ reddening map of \citet[SFD98]{sfd98}} \\
\hline
BDF\_T & $\asec^2$ & Intrinsic squared size of best-fit BDF model, before PSF convolution: $T = \langle x^2 \rangle  + \langle y^2 \rangle$\\
\hline
BDF\_T\_ERR & $\asec^2$ & Estimate of error in BDF\_T\\
\hline
BDF\_T\_RATIO & & Ratio of BDF\_T of the object to PSF\_T at the location of the object (stars are near zero). \\
\hline
BDF\_FRACDEV & & Fraction of light in a bulge (Sersic $n=4$ model)\\
\hline
BDF\_G\_(1/2) & & BDF ellipticity components \\
\hline
EXT\_(COADD/FITVD/MASH/WAVG/XGB) & & \makecell[cl]{Classification code for the `extendedness' of object, \\from 0 (point-like) to 4 (extended-like). See Section \ref{sec:classification}.}\\
\hline
XGB\_PRED & & \makecell[cl]{Predictor output from the XGBoost star/galaxy classifer.
\\ Galaxies have XGB\_PRED $\sim 0$ and stars have XGB\_PRED $\sim 1$. \\ See Section \ref{sec:classification}.}\\ 
\hline
FLAGS\_FOOTPRINT & & Flag indicating that the object belongs to \gold. See Section \ref{sec:footprint}.\\
\hline
FLAGS\_GOLD & & Flag showing possible processing issues with the object. See Section \ref{sec:flagsgold}\\
\hline
FLAGS\_FOREGROUND & & \makecell[cl]{Flag showing that the object is in the area of influence of \\ a foreground object from an imaging point of view. See Section \ref{sec:foregrounds}.}\\
\hline
DNF\_(Z/ZN) & & DNF photo-z estimate for the object, using DNF\_NNEIGHBORS or the nearest neighbor. See Section \ref{sec:photoz}.\\ 
\hline
DNF\_ZSIGMA & & \makecell[cl]{DNF photo-z uncertainty estimate from photometric uncertainties and residuals \\ from the neighborhood fit. }\\
\hline
DNF\_NNEIGHBORS & & Number of neighbors used for the DNF\_Z estimate \\
\hline
DNF\_ZERR\_PARAM & & The uncertainty on DNF\_Z due to photometric errors \\
DNF\_ZERR\_FIT & & The uncertainty on DNF\_Z from the residuals of the fit \\
\hline
\enddata
\tablecomments{
Names in parentheses show options for a given type of column separated by slashes for each column. 
Full details at \drurl.
}

\end{deluxetable}

In \tabref{catalog_columns} we summarize the essential columns of the \gold data set with their brief description. 
Full details will be provided upon release at \drurl.

\section{Survey Property Maps}
\label{app:survey_properties}

Survey property maps are computed from a base  \mangle polygon file and converted to \healpix maps as described in Appendix E of \citet{Y3Gold}.

In addition to these, certain survey property maps were also created using \code{decasu} which is meant to be a complete replacement for \mangle mapping, running a high resolution pixelized map quickly and efficiently. This sofware natively uses \healsparse formatted maps that are designed to store high resolution information without dramatically increasing the memory usage (therefore allowing to go beyond the limit of \nside = 4096 imposed by standard RAM machine limitations). While \mangle is better at describing the maps at the highest ("true") resolution, in practice we cannot make use of these maps without pixelizing and degrading them.

In \tabref{observingconditions} we summarize the observing conditions per band. 

\begin{figure}[ht]
\centering
\includegraphics[width=0.45\textwidth]{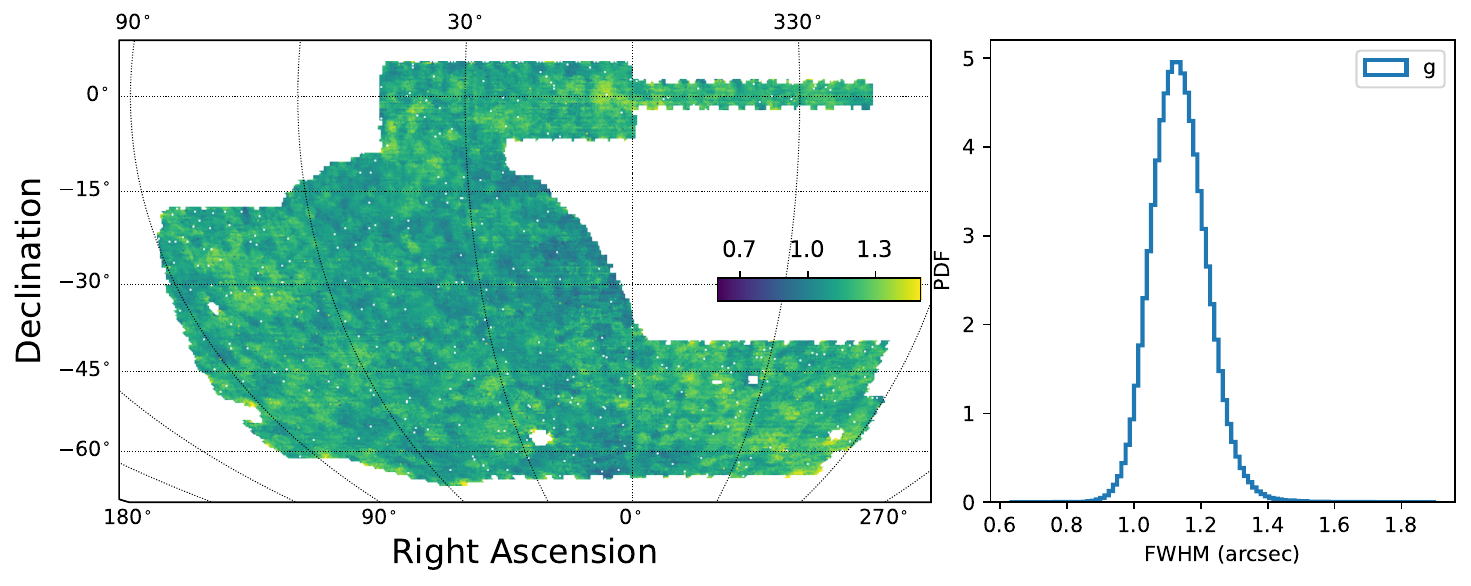}
\includegraphics[width=0.45\textwidth]{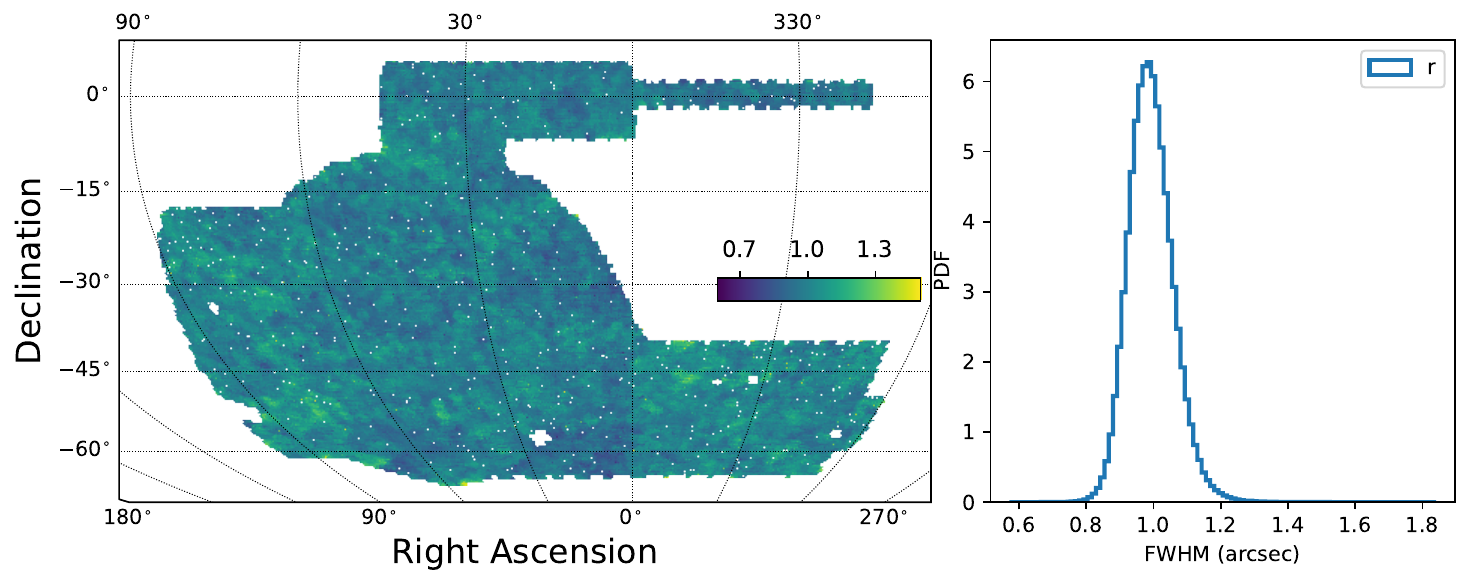}
\includegraphics[width=0.45\textwidth]{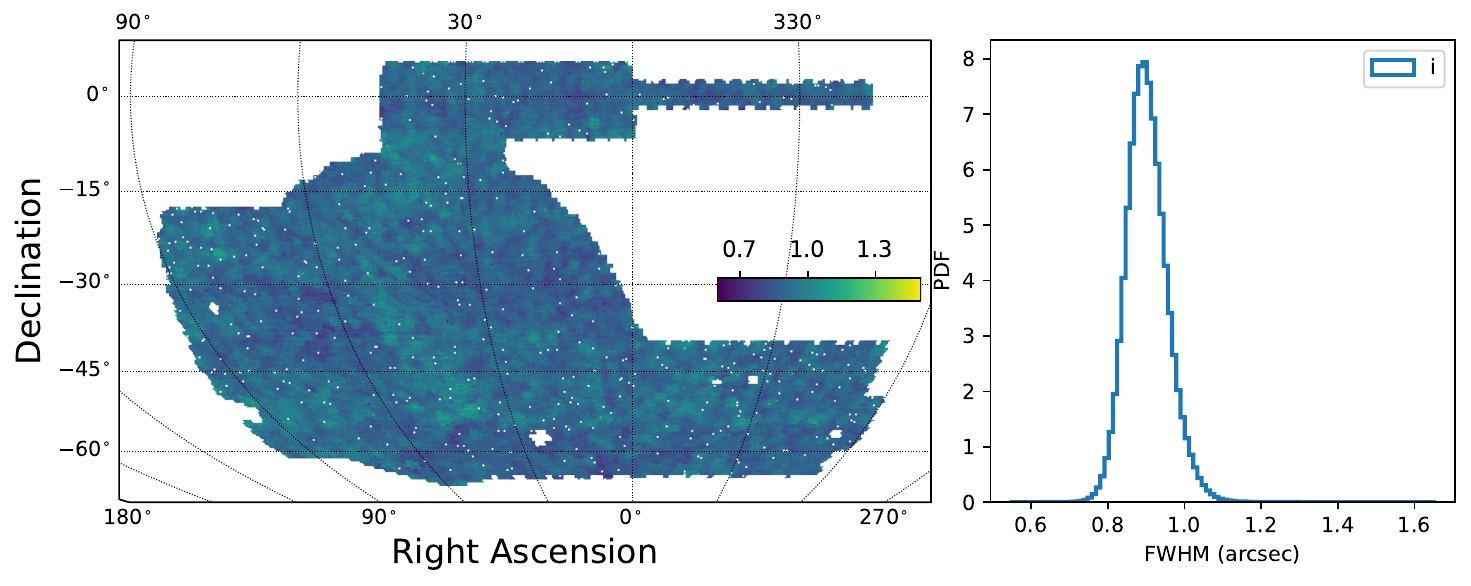}
\includegraphics[width=0.45\textwidth]{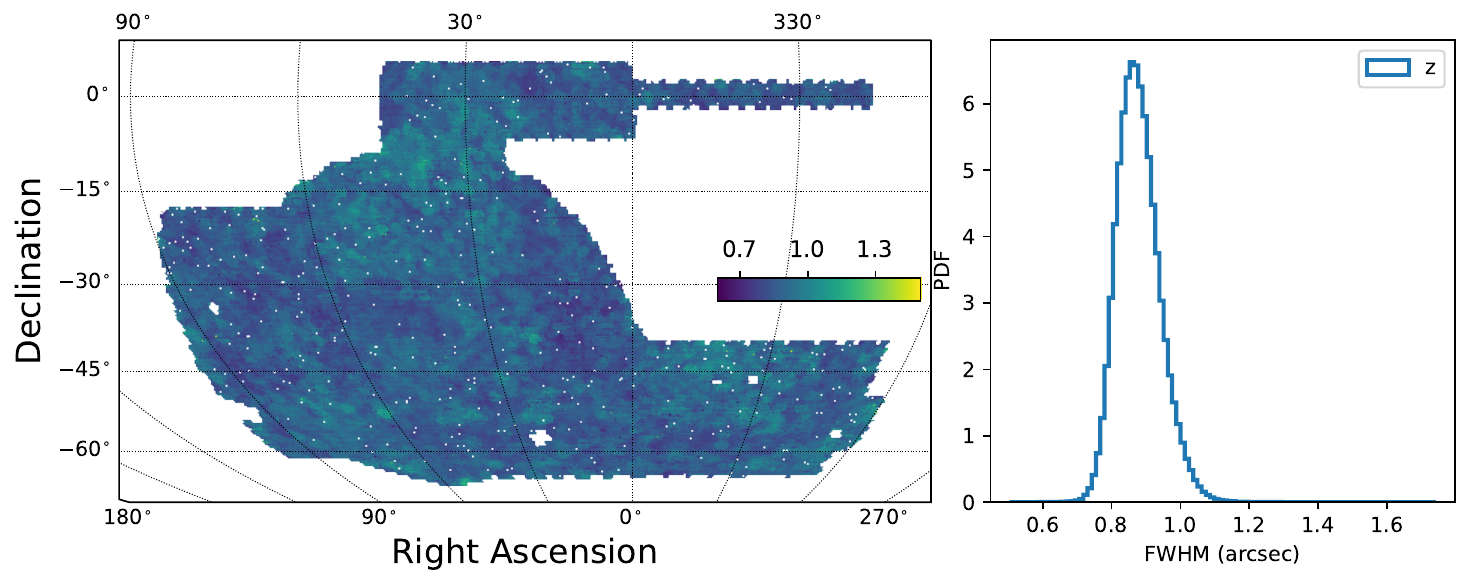}
\includegraphics[width=0.45\textwidth]{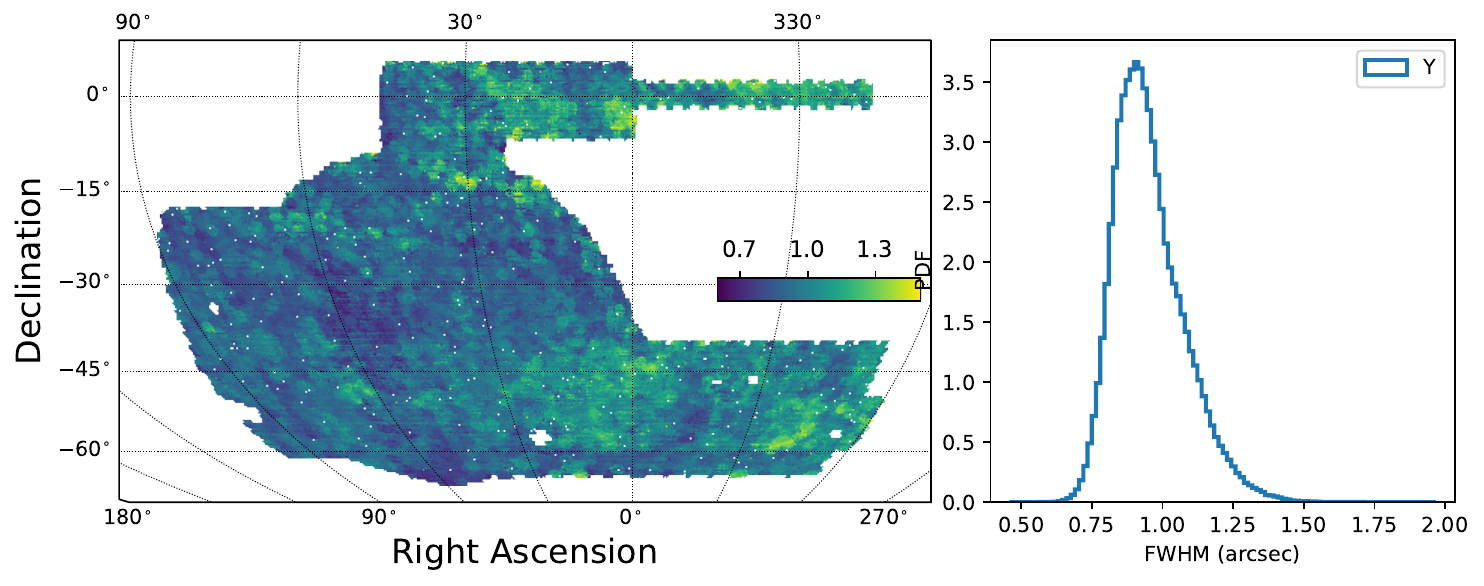}
\caption{\label{fig:app_fwhm} 
Sky maps and histograms of the seeing (fwhm\_wmean) for each of the observed bands. The value at each location is the inverse-sky-variance-weighted sum of all individual exposures of that \healpix pixel.}
\end{figure}

\begin{figure}[ht]
\centering
\includegraphics[width=0.45\textwidth]{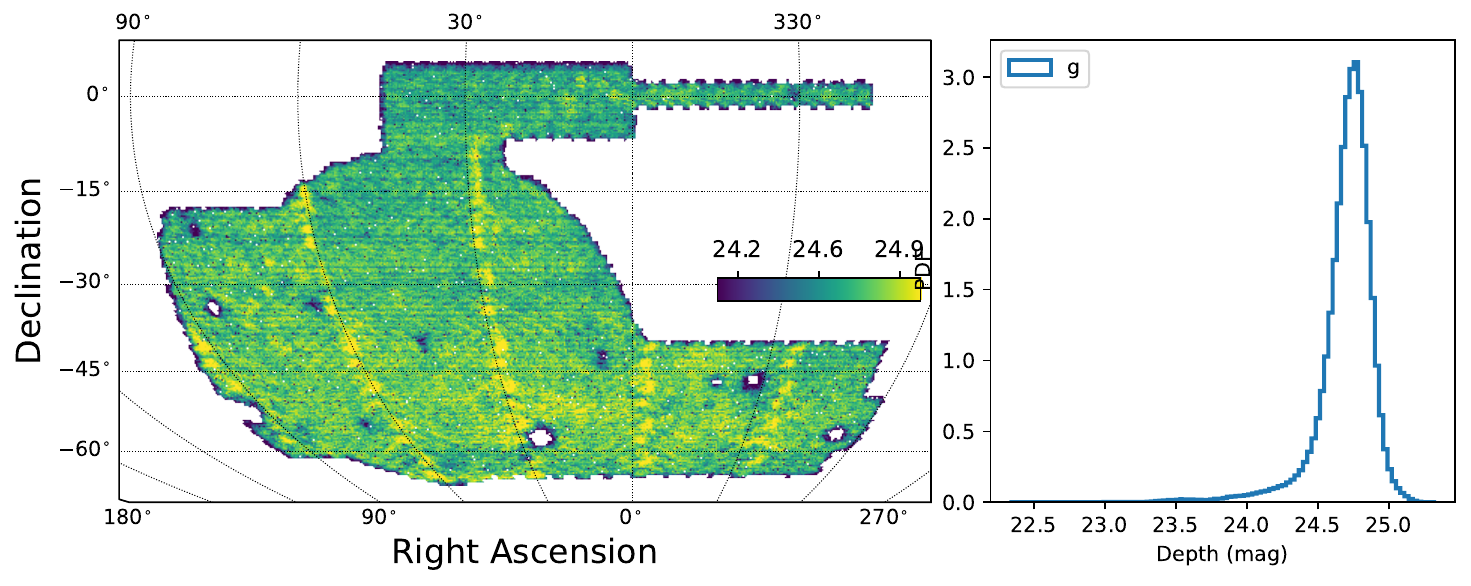}
\includegraphics[width=0.45\textwidth]{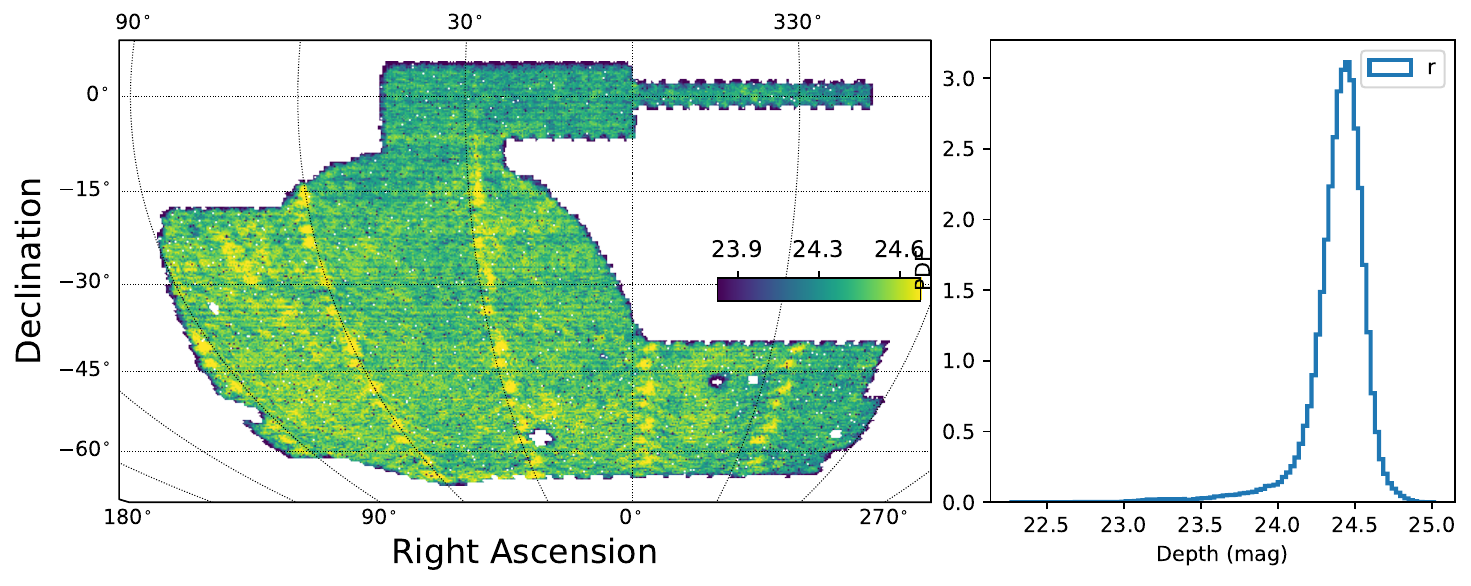}
\includegraphics[width=0.45\textwidth]{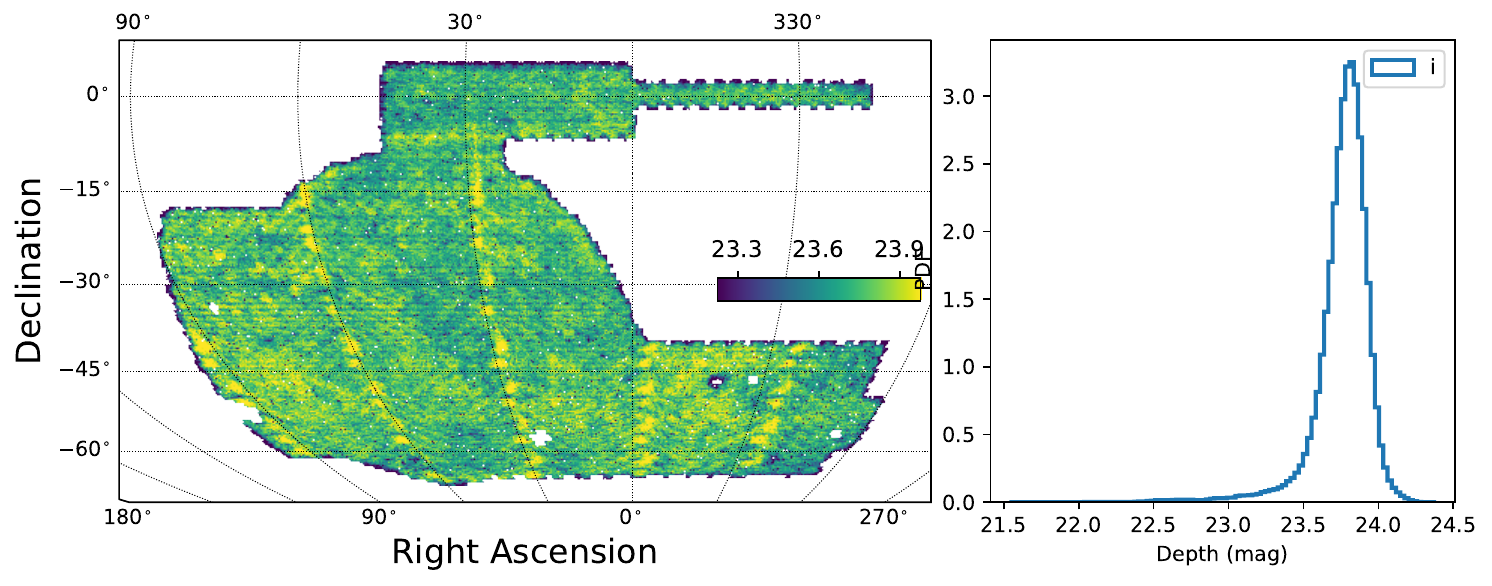}
\includegraphics[width=0.45\textwidth]{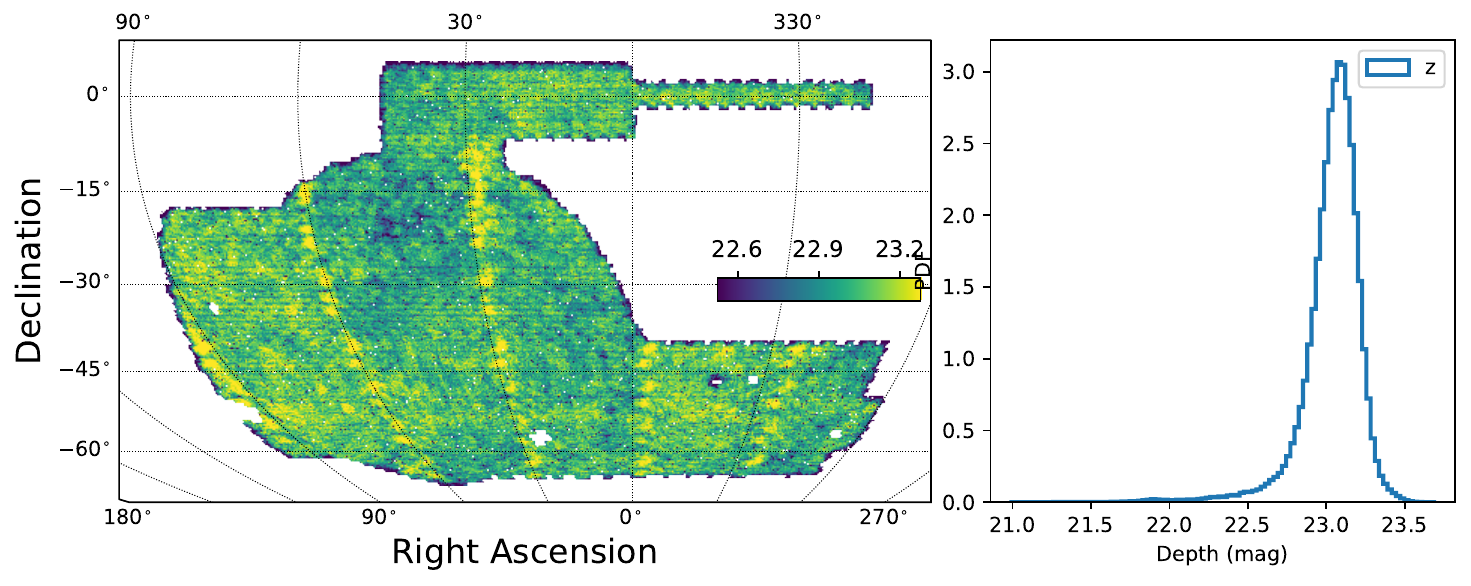}
\includegraphics[width=0.45\textwidth]{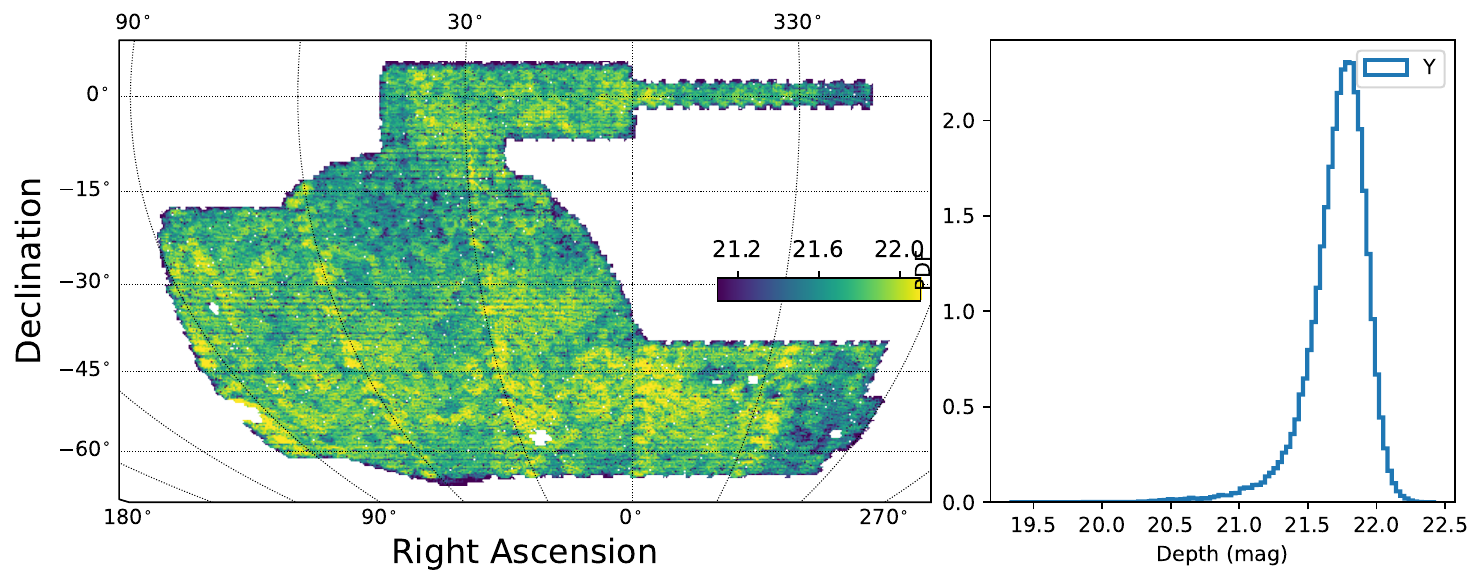}
\caption{\label{fig:app_maglim} 
Sky maps and histograms for the magnitude limit (maglim\_wmean) estimated from the weight maps.  Note that the linear features along equal RA values are a result from regions covered by more than 10 exposures per band, which are inevitable when attempting to tile the sphere with no less than 10 exposures per location.}
\end{figure}

\figref{app_fwhm} and \figref{app_maglim} show two example maps as a function of position in the sky and the corresponding histogram of computed values for these positions (computed in \nside = 4096 \healpix resolution). Note that the linear features along equal RA values are a consequence of the observation strategy to ensure a complete tiling of the sphere.

\subsection{Cirrus / Nebulosity Maps}
\label{app:cirrus}

Although the DES footprint avoids the Galactic plane, some regions are nonetheless affected by Galactic cirrus/nebulosity. These are being incorporated for the first time to the Gold suite of survey property maps with \gold.

Galactic cirrus manifests as faint diffuse light with surface brightness variations on scales between a few arcseconds and a few arcminutes.  
For comparison, background estimation using the AstrOmatic software (both \sextractor and \swarp) sampled a scale of 256$\times$256 pixels ($\sim$67\asec$\times$67\asec) and therefore, unaccounted structured diffuse light on smaller angular scales can impact object detection and measurements.  
To better understand the extent of cirrus light, and potentially provide a means to quantify or mitigate its impact on source detection and measurement, we investigated the use of an existing machine learning application, \code{MaxiMask}\footnote{This work used \url{https://github.com/mpaillassa/MaxiMask/} version 1.0.} \citep{2020A&A...634A..48P}, that was trained on individual DECam observations to identify a number of different characteristics, with nebular/diffuse light being the one of interest here. 
For faint diffuse light, we find that the default training provides a good discriminant for the presence of this nebulosity for the DES Y6 coadded images, and even spatially binned coadded images.  

To obtain a map of nebular light across the DES footprint, we performed the following steps for each of the 5 DES bands for every DES coadd tile:
\begin{enumerate}
    \item Start with the DES \code{coadd\_nobkg} image products that were assembled by \swarp without applying a background subtraction (i.e., \var{-SUBTRACT\_BACK N}).
    \item Bin each coadd image by calculating the median of  the values in groups of 5$\times$5 pixels. 
    \item Run \code{MaxiMask} to obtain a probability-like estimate, $\zeta$, that the flux detected in each binned pixel is consistent with a diffuse/nebular origin.
    \item Map each binned pixel onto \healpix grids with $\nside=$1024 and 4096 (NESTED) and then accumulate statistics within each \healpix element to form maps of median and maximum values of $\zeta$, as well as the median and maximum surface brightness.
\end{enumerate}

The resulting maps of max($\zeta$) in the $gri$-bands show good correspondence to maps of extinction and to H{\sc I} surveys when constrained to high velocities (i.e., Galactic cirrus).  
At longer wavelengths ($zY$-bands) the detection of diffuse nebulosity is less significant and the correspondance to the extinction and H{\sc I} maps is less pronounced.

\begin{deluxetable}{c c l}[h]
\tablewidth{0pt}
\tabletypesize{\tablesize}
\tablecaption{ \gold Survey Properties.  \label{tab:observingconditions}}
\tablehead{
\colhead{\textbf{DES map name (from \mangle)}\tablenotemark{\tiny a}} & \colhead{Units} &
\colhead{Description} 
}
\startdata
NUMIMAGE & & Number of images \\
\hline
MAGLIM & & Magnitude limit estimated from the weight maps \tablenotemark{\tiny c} \\
\hline
FRACDET & & Effective area fraction considering the bleed-trail and bright star masks \\
\hline
EXPTIME.SUM &  seconds & Exposure time \\
\hline
T\_EFF.(WMEAN/MAX/MIN) & & Figure of merit for quality of observations $t_{eff}$ \tablenotemark{\tiny d}\\
\hline
T\_EFF\_EXPTIME.SUM &  seconds &  Exposure time multiplied by $t_{eff}$\\
\hline
SKYBRITE.WMEAN & electrons/CCD pixel & Sky brightness from the sky background model \tablenotemark{\tiny e} \\
\hline
SKYVAR.(WMEAN/MIN/MAX) & (electrons/CCD pixel)$^2$ & Variance on the sky brightness \tablenotemark{\tiny f} \\
\hline
SKYVAR\_SQRT.WMEAN & electrons/CCD pixel & Square root of sky variance \\
\hline
SKYVAR\_UNCERTAINTY & electrons/s/coadd pixel & Sky variance with flux scaled by zero point \\
\hline
SIGMA\_MAG\_ZERO.QSUM & mag & Quadrature sum of zeropoint uncertainties. \\
\hline
FWHM.(WMEAN/MIN/MAX) & \asec  & Average FWHM of the 2D elliptical Moffat function that fits best the PSF model from \code{PSFEx} \\
\hline
FWHM\_FLUXRAD.(WMEAN/MIN/MAX) & \asec & Twice the average half-light radius from the sources used for determining the PSF with \code{PSFEx} \\
\hline
FGCM\_GRY.(WMEAN/MIN/MAX) & mag & Residual `gray' corrections to the zeropoint from FGCM \\
\hline
AIRMASS.(WMEAN/MIN/MAX) & & Secant of the zenith angle \\
\hline
CIRRUS\_NEB\_(MEAN/MAX) & & Mean or maximum probability of nebular emission \\
\hline
CIRRUS\_SB\_(MEAN/MAX) & & Mean surface brightness in image pixels (relative values)  \\
\hline
\hline
\textbf{DES map name (from \decasu)}\tablenotemark{\tiny b} & Units & \hspace{5.5cm} Description \\
\tableline
airmass\_wmean &  & Secant of the zenith angle\\
\hline
fwhm\_wmean & pixels & Average FWHM of the 2D elliptical Moffat function fit to the \code{PSFEx} model (0.263 \asec/pix)\\
\hline
maglim\_wmean & mag & Magnitude limit estimated from the weight maps\\
\hline
nexp\_sum  & & Number of exposures \\
\hline
exptime\_sum & seconds & Exposure time\\
\hline
skybrite\_wmean(\_scaled)  \tablenotemark{\tiny g} & electrons/CCD pixel & Sky brightness from the sky background model\\
\hline
skysigma\_wmean(\_scaled)  \tablenotemark{\tiny g} & electrons/CCD pixel & Square root of sky variance \\
\hline
dcr\_(dra/ddec/e1/e2)\_wmean &  & Differential chromatic refraction effect on positions and ellipticity (relative shifts)\\
\hline
bdf\_nside(4096/16384)(\_nodered)\_depth & mag & Magnitude limit, using raw or dereddened magnitudes, in \nside = 4096/16384 \healpix resolution \\
\hline
\enddata
\tablecomments{
Survey properties in \gold registered as maps. Each quantity has been calculated individually for $grizY$ bands. 
\tablenotetext{a} {These maps are produced in \healpix format in $\nside=4096$ in \code{NESTED} ordering, averaging from a higher resolution version ($\nside=32768$). Each high resolution pixel adopts the value of the molygon from the \mangle map at its center, which is a statistic of a stack of images contributing to that point in the sky.  \code{WMEAN} quantities are the mean value weighted using the weights obtained from \mangle. \code{MIN, MAX} correspond to the minimum or maximum of all the stacked images in the molygon. \code{SUM} adds up the contribution of all images to the molygon. \code{QSUM} makes a quadrature sum instead. The DES map name is the name given to the files as they are delivered in the release page.}
\tablenotetext{b}{These maps are produced in \healsparse format, with weighted means unless indicated otherwise.}
\tablenotetext{c}{10-$\sigma$ magnitude limit in 2\asec diameter apertures.}
\tablenotetext{d}{$t_{eff}$, as described in \citet{desdm}, Equation 4, is measured as a ratio between exposure time and the exposure time necessary to achieve the same signal-to-noise for point sources observed in nominal conditions. This depends on a set of fiducial conditions per band for full-width half maximum, sky background and atmospheric transmission.}
\tablenotetext{e}{The model value used is taken as the median per CCD. Details for this model are described in \citet{Bernstein:2017} and \citet{desdm}.}
\tablenotetext{f}{Takes into account intrinsic sky Poisson noise, read noise and flat field variance.}
\tablenotetext{g}{Scaled quantities indicate that the electron/pixel values have been scaled according to the zeropoints of the coadds.}
}

\end{deluxetable}

\bibliographystyle{yahapj_twoauthor_arxiv_amp}
\bibliography{biblio}

\end{document}